\documentclass[12pt,preprint]{aastex}

\usepackage{natbib}

\shorttitle{AMiBA Correlator}
\shortauthors{Li et al.}

\begin{document}


\title{AMiBA Wideband Analog Correlator}


\author{Chao-Te Li\altaffilmark{1}, Derek Y. Kubo\altaffilmark{1}, Warwick Wilson\altaffilmark{2}, Kai-Yang Lin\altaffilmark{1,3}, Ming-Tang Chen\altaffilmark{1}, P. T. P. Ho\altaffilmark{1,4}, Chung-Cheng Chen\altaffilmark{1}, Chih-Chiang Han\altaffilmark{1}, Peter Oshiro\altaffilmark{1},
Pierre Martin-Cocher\altaffilmark{1}, Chia-Hao Chang\altaffilmark{1}, Shu-Hao Chang\altaffilmark{1}, Pablo Altamirano\altaffilmark{1},
Homin Jiang\altaffilmark{1}, Tzi-Dar Chiueh\altaffilmark{3}, Chun-Hsien Lien\altaffilmark{3}, Huei Wang\altaffilmark{3}, Ray-Ming Wei\altaffilmark{3}, Chia-Hsiang Yang\altaffilmark{3}, Jeffrey B. Peterson\altaffilmark{5}, Su-Wei Chang\altaffilmark{1}, Yau-De Huang\altaffilmark{1}, Yuh-Jing Hwang\altaffilmark{1}, Michael Kesteven\altaffilmark{2}, Patrick Koch\altaffilmark{1}, Guo-Chin Liu\altaffilmark{1,6}, Hiroaki Nishioka\altaffilmark{1}, Keiichi Umetsu\altaffilmark{1}, Tashun Wei\altaffilmark{1} and Jiun-Huei Proty Wu\altaffilmark{3}}

\altaffiltext{1}{Institute of Astronomy and Astrophysics, Academia Sinica, P.O. Box 23-141, Taipei, Taiwan 106}
\altaffiltext{2}{Australia Telescope National Facility, Epping, NSW Australia 1710}
\altaffiltext{3}{National Taiwan University, Taipei, Taiwan 106}
\altaffiltext{4}{Harvard-Smithsonian Center for Astrophysics, 60 Garden Street, Cambridge, MA 02138, USA}
\altaffiltext{5}{Carnegie-Mellon University, Pittsburgh, PA 15213 USA }
\altaffiltext{6}{Tamkang University, 251-37 Tamsui, Taipei County, Taiwan}

\email{ctli@asiaa.sinica.edu.tw}


\begin{abstract}
A wideband analog correlator has been constructed for the Yuan-Tseh Lee Array for Microwave Background Anisotropy. Lag correlators using analog multipliers provide large bandwidth and moderate frequency resolution. Broadband IF distribution, backend signal processing and control are described.  Operating conditions for optimum sensitivity and linearity are discussed. From observations, a large effective bandwidth of around 10 GHz has been shown to
provide sufficient sensitivity for detecting cosmic microwave background variations.

\end{abstract}




\section{Introduction} 
Interferometric observations have gained much popularity in the study of the cosmic microwave background (CMB) anisotropy (White et al. 1999; Carlstrom et al. 2002; Padin et al. 2002; Leitch et al. 2002; Scaife et al. 2008), due to their advantage in stability and phase preserving characteristic via 
the heterodyne scheme for correlation and polarization observations. The cross correlations used in an interferometric array can effectively suppress many 
systematics. To achieve high brightness sensitivity, broadband low noise receivers and matching correlators are the two most important components for 
a continuum interferometer. The advance in millimeter and microwave detector technologies in recent years has produced very broadband components. Low-noise amplifiers (LNAs) with bandwidths of 10 GHz or more are easily accessible with noise performance comparable to bolometric direct detectors, e.g. ACBAR (Runyan et al. 2003) with bandwidths of 30 - 40 GHz and an equivalent noise temperature or RJ temperature loading $T_{RJ}$ between 40 K and 70 K. The CMB interferometers are therefore very competitive at millimeter wavelengths as compared to bolometers which are usually preferred at shorter wavelengths. 

The Yuan-Tseh Lee Array for Microwave Background Anisotropy (AMiBA) (Lo et al. 2001; Ho et al. 2009) is a radio interferometric array for the study of the CMB at 3mm wavelength. AMiBA detects the minute deviations of the nearly constant CMB temperature over the sky, and can study the spatial variation of this temperature fluctuation. In particular, AMiBA is imaging galaxy clusters via the Sunyaev-Zel'dovich effect (SZE) (Sunyaev and Zel'dovich 1970, 1972;
Birkinshaw 1999) for the first time at 3 mm wavelength. The array presently consists of 13 elements of 1.2 m reflectors distributed on a 6-m diameter platform. The receiver attached to each reflector is dual-polarization, and equipped with cryogenically cooled LNAs operating in the 84 to 104 GHz frequency range. The intermediate frequency (IF) is from 2 to 18 GHz, and is matched with a wide-band analog correlator. 

The strong interest in CMB observations has motivated the development of very broadband correlators with a limited spectral resolution. Utilizing a filter bank and complex correlators, Padin et al (2002) built an analog correlator with 10 GHz bandwidth. Harris and Zmuidzinas (2001) worked on a different approach toward broadband operations by adopting an analog lag correlation scheme to build an auto correlator with a 4 GHz bandwidth. The AMiBA correlator is also based on the concept of the lag correlator, and is designed to correlate the entire 16 GHz bandwidth. Recently Holler et al. (2007) also reported a lag correlator for AMI with a 6 GHz bandwidth. 

In principle, analog correlators can achieve better sensitivity over their digital counterparts due to the absence of the digitization process. 
Analog multipliers can easily achieve high sensitivity over multi-octave frequency ranges. In comparison the use of analog-to-digital converters (ADCs) is expensive and has limited bandwidth. Thus the analog approach is preferred for observations which require high sensitivity but modest spectral resolution. However, the major difficulty in making a broadband system lies in the distribution and processing of the multi-octave signals coming from the antennas. Broadband impedance matching between components presents a major technical challenge for integrating a large scale microwave system. Due to the non-linear responses of an analog system, applying appropriate drive power levels and modulation/demodulation techniques to minimize the effects of the spurious terms are also very important. The AMiBA correlator is our attempt to address these technical issues. The 4-lag analog correlator has a nominal 16 GHz bandwidth. This is currently the only correlator in operation with an effective correlation bandwidth of around 10 GHz.

The scientific goals and design philosophy of AMiBA are presented in Ho et al. (2009). A broader description of the AMiBA detection system is given
in Chen et al. (2009) while a detailed description of the AMiBA hexapod mount can be found in Koch et al. (2009). From 2007 to 2008, observations were carried out with the 7-element array equipped with 60 cm dishes (Koch et al. 2006). Details on the observations and analysis of six massive
galaxy clusters are presented in Wu et al. (2009). Subaru weak lensing data of 4 galaxy clusters were analyzed with the SZE data to derive
the baryon fraction (Umetsu et al. 2009). This paper describes in detail the instrumental design and testing of the AMiBA correlator. We provide a system overview in Section 2 and major components of the system are described in Sections 3 to 7. In particular, we discuss various aspects of the correlator module in Section 4. In Section 8, we outline the testing and data processing. Finally, a conclusion on the system is given in Section 9.

\section{System Overview}

As shown in Fig. 1, the AMiBA correlator consists of 5 parts, namely the IF distribution, correlation, readout, data acquisition, and control. For the 13-element array, there would be a total of $(13 \times 12/2)\times 4 = 312$ correlations between the 2 polarizations of each receiver in order to obtain the 4 Stokes parameters for polarization measurements. In AMiBA correlator, the number of correlator modules required is reduced by a factor of two by introducing 2-way switches which choose between the two parallel-hand products, XX and YY, or the cross polarization products, XY and YX. For example, observations of the CMB spatial intensity variations require measurement of the XX and YY products only.

The network to distribute the 2 to 18 GHz IF from the receivers to the correlators is implemented by cascading 3 sections of 4-way power dividers. The broadband analog lag correlator modules were designed using double-balanced diode mixers as the multipliers. The subsequent amplifier provides low-pass
filtering with a 3 dB cut-off frequency at around 10 KHz. 

In the readout electronics, we use voltage-to-frequency converters (VFCs) and counters to provide integrating analogue to digital conversion of the multiplier outputs. This style of ADC has a slow response but a high dynamic range, making them well suited to this application. The data acquisition electronics serve as a memory buffer between the readout electronics and the correlator control computer (CCC). In response to timed event signals from the CCC, it also generates control signals for phase switching, demodulation, and the readout process.

The CCC coordinates all the activities in the correlator, as well as archiving the data. Equipped with 3 special purpose cards from the Australia Telescope National Facility (ATNF), the CCC provides timing, data interface, and event signals for correlator operations. Further processing of the archived data is performed offline. This includes the processing required to transform the four lag domain measurements from each lag correlator into two complex channels in the frequency domain. In a digital lag correlator, the conversion from the lag domain to the frequency domain is a simple discrete Fourier transform. In the analogue lag correlator, variations in gain and bandpass of the individual multipliers complicate the transform process considerably. Calibration
with strong point sources such as planets becomes necessary. By observing a strong point source, the expected signal from each frequency bin of each baseline can be calculated. Correction to the transformation can then be extracted and applied to the following observations. A brief description of the relevant data processing and demonstration are presented in Section 8.

\section{IF distribution}  
An IF from 1 GHz to 21 GHz was proposed in the beginning of the project (Lo et al. 2001), but was subsequently changed to 2 GHz to 18 GHz after taking into consideration the available commercial microwave components\footnote{Commercial broadband microwave components usually come with a frequency range up to 18 GHz, or up to 26 GHz for the next available frequency range.}, the complexity of the multi-octave circuit design, and the physical sizes of components. The IF signal distribution uses a mixture of off-the-shelf and specially designed microwave components.

After passing through the RF LNAs and 
sub-harmonically pumped mixers (SHMs), which have a conversion loss of around 12 dB, the faint signals from the receiver inputs reach a level of -45 to
-40 dBm. At this point, the noise figure of the subsequent components has a negligible contribution to the overall noise temperature, as long as
the IF power level can be kept well above the thermal noise of about - 71 dBm. Multiple broadband IF amplifiers are used along the IF chain to compensate for the divider loss. Attenuators are inserted along the IF paths to adjust the IF power level in order not to saturate the amplifiers, and also to improve the matching between components. The input power level to the correlator modules is chosen to optimize the output signal-to-noise (S/N) ratio. The diagram of IF power level settings from receiver to correlator is shown in Fig. 2.

Another issue is to keep the IF power stable over a period greater than the integration time of each data point. It was found that most of the gain variations can be attributed to ambient temperature changes (Nishioka et al. 2009). A ventilation system using fans and a feedback proportional-integral-derivative (PID) control was installed to minimize the temperature variations within the electronic cabinets (Jiang et al. 2009).

The following sections provide a description of each IF section. Schematics of the AMiBA receiver and correlator IF sections are shown in Figs. 3 and 4 respectively for reference.

\subsection{$1^{st}$ Section}
In the $1^{st}$ section, an IF amplifier\footnote{CMA-18-2004, Teledyne Microwave} with a gain of 34 dB and output 1dB compression point (P1dB) of 20 dBm is used in conjunction with a 4-way power divider\footnote{PDM-44M-10G, Merrimac}. A directional coupler\footnote{1822, Krytar} sends part of the
IF signal into a total power detector\footnote{302A, Krytar} to monitor the IF power after the receiver. The reading can be used to adjust the variable
gain amplifier\footnote{AVG4-02001800-40, Miteq} (VGA) in the receiver IF to maintain the power level from the receiver. In addition to monitoring the receiver gain variations, the total power reading can also be used for sky dip or hot/cold load experiments to determine the sky and receiver noise temperature. The input of the total power readout electronics is switched between the signal and ground for alternating data. Common mode noise such as ground noise\footnote{As the total power readout electronics input is connected to ground through a 1M ohm resistor, the fluctuations on the reading can be seen. Similar effect can be observed as the readout is connected to the total power signal while the input of the $1^{st}$ section is either terminated with a 50 ohm resistor or connected to a stable IF signal.} are reduced after subtracting the ground reference from the signal.

An 18 GHz lowpass filter\footnote{11SL10-1800/X24000-O/O, K\&L Microwave} determines the overall system bandwidth and also filters out the 21 GHz LO leakage from the mixer. A 13-dB negative slope equalizer\footnote{EQ1251-13M, Aeroflex/Inmet} compensates for the gain slope arising from the following components.

\subsection{$2^{nd}$ Section}
There are two versions of the $2^{nd}$ section - one with a 2-way switch\footnote{521-420803A, Dow-Key Microwave}, a delay trimmer\footnote{981, Aeroflex/Weinshel}, an IF amplifier, and a 4-way power divider; the other with an IF amplifier, a power divider, and a delay compensation cable. During polarization observations, the 2-way switches can be utilized to obtain 4 cross correlations. The delay trimmer has a range of +/- 90 pico-seconds for fine delay tuning. The coarse delay adjustment is done by installing delay compensation cables of designated lengths. More details about delay trimming can be found in Lin et al. (2009).

\subsection{$3^{rd}$ Section}
The $3^{rd}$ section consists of a 4-way power divider with built-in amplifiers in front. Their dimensions are custom designed in order
to feed the wideband IF signals into an array of correlator modules in a compact way as shown in Fig. 5. The two stages of the MMIC IF amplifiers\footnote{TGA8300-SCC, Triquint Semiconductor} with a cascaded gain of 14 dB are placed in front of the power divider to compensate for the loss in the power divider. A millimeter-wave DC-blocking capacitor\footnote{500S100GT 50 XT, American Technical Ceramics} required for bias is placed in front of the $1^{st}$ MMIC and rejects signals below 0.5 GHz. The power divider utilizes a Cu-clad substrate with an embedded resistive layer\footnote{Arlon 25N + OhmegaPly} which allows thin-film resistors to be etched into the circuit as part of the fabrication process. However, the same resistive layer causes significant loss and gain slope along a long transmission line. For cavity resonance suppression, metallic irises were extended down from the cover to confine the circuits into small channels. A picture of the custom power divider is shown in Fig. 6. The component layout of the amplification stage and layouts of the two types of power dividers are also shown in Fig. 7 and 8, respectively, for illustration. Details on the circuit design and test results can be found in Li et al. (2004).

\section{Correlator}
\subsection{Design and Specifications}
In a lag correlator system as for AMiBA, a number of multipliers are used to measure the correlation as a function of the time offset or "lag" between two signals, namely 
\begin{equation}
r(\tau) = V_1(t) \star V_2(t) = \lim_{T\to\infty} 
\frac{1}{2T} \int^T_{-T} V_1(t)V^*_2(t-\tau) dt,
\end{equation}
where $\tau$ is the lag between the two signals. The cross correlation is represented by the pentagram symbol ($\star$), and $V^*_2$ denotes the complex conjugate of $V_2$. The cross power spectrum of $V_1(t)$ and $V_2(t)$ can be derived via Fourier transforming the cross correlation
\begin{equation} 
V_1(t) \star V_2(t) \rightleftharpoons \hat{V}_1(\nu) \hat{V}^*_2(\nu).
\end{equation}
In a lag correlator of the type described here, the correlation function is measured at discrete values of lags. The bandwidth of the correlator, $BW$, is determined by the delay increment $\delta\tau$ between measurements according to the Nyquist sampling theorem (Harris et al. 2001). The frequency resolution, $\delta f$, is determined by the number of lag measurements, $N$, such that $\delta f = BW/(N/2)$.

To limit the gain loss from bandwidth smearing to an acceptable level, for an efficiency of $sin(x)/x = 0.9$ at the primary
beam half power points, the fractional bandwidth must be at most
\begin{equation}
\frac{\delta\nu}{\nu} = \frac{1}{2}\frac{D}{B},
\end{equation}
where $B$ is the baseline length and $D$ is the dish diameter. Given the final configuration of the array (1.2m dishes with longest baselines of about 6 m), a frequency resolution of 8 GHz is chosen. With a bandwidth of 16 GHz, the number of lags in the lag correlator is then set to 4 to provide 2 frequency channels. Correlator modules with more (e.g. 8) lags would not furnish much more information but would minimize the bandwidth smearing for the case of smaller diameter dishes. To examine the bandwidth smearing from another persepctive, Fig. 9 shows the calculated bandwidth patterns of the 4 lag outputs
with an effective bandwidth of 16 GHz. The lag span is between +/- 60 ps, enough to cover the delay for 1.2m dishes with 6 m baselines. 

For the AMiBA correlator, it is important to have a very wide bandwidth for CMB observations. To obtain the wide bandwidth, analog multipliers in the form of balanced mixers were employed. The use of passive multipliers circumvents the problem of 1/f noise and other noise from the bias circuitry usually associated with active multipliers. A flat amplitude response and a linear phase response (non-dispersive) are essential to achieve a large effective bandwidth. From tests with a translating noise source, the entire signal path with the 4-lag correlator module has an effective bandwidth of around 10 GHz (Lin et al. 2009). A brief list of the AMiBA correlator specifications is given in Table 1. 

Within each correlator module, two stages of 2-way power dividers
are cascaded to split the IF signals to feed four multipliers. 
The lags for each multiplier are specified to be 37.5, 12.5, -12.5,
and -37.5 pico-seconds, respectively. 
The 25 pico-second lag spacing is designed to provide Nyquist sampling
of the IF up to 20 GHz. The output voltages from the 4 lags are transformed 
into 2 complex data points in the cross power spectrum. The double-balanced mixers
comprise of low-barrier Silicon Schottky diode ring quads mounted between two wideband baluns.
To avoid reflections due to discontinuities such as wire bonding between
circuits, the entire circuit, including power dividers and baluns,
is manufactured on a single microwave substrate.
A picture of the 4-lag correlator module manufactured by Marki Microwave Inc. 
is shown in Fig. 10. 

Regarding the channel isolation, for an ideal 4-lag correlator, the recovered spectrum is 
$\Sigma_i c_i e^{- j \omega \tau_i},$
where $c_i$ is the correlation measurement at each lag, and $\tau_i$ is the nominal lag. If the input signals are continuous-wave (CW) signals at the center of each channel, namely 6 GHz and 14 GHz in our case, the recovered spectrum is shown in Fig. 9 with a FWHM of about 12 GHz for each CW signal. Therefore, the leakage between channels is quite severe for the 4-lag correlators. For comparison, the recovered spectrum of an ideal 8-lag correlator with input CW signals at the center of its 4 channels is also shown. Although the leakage between adjacent channels is still severe for an 8-lag module, the isolation between non-adjacent channels is acceptable. To improve on the channel isolation, one possibility is to apply a window function during the lag-to-visibility transformation. However, constrained by the small number of lags that we have, another possibility might be to apply bandpass filtering or a filter bank before correlation. The filter bank might also reduce the bandpass variation within each channel and maximize the effective bandwidth.

\subsection{Optimum Input Power}
During observations, the correlator input signals consist of large un-correlated 
noise signals from the receivers and tiny correlated signals from the sky.  
These large un-correlated noise signals tend to pump the correlator diodes
in much the same way as the local oscillator signal in a mixer application.
This can lead to non-linearity of the correlator multipliers and an excess
of noise. We adjust the correlator input power level to reduce
this degradation.

In a simulation of a double-balanced mixer with four tones
(two with large power and two with 28 dB smaller power)
as shown in Fig. 11, when
the large signals have a power below a certain level, the
product of the two small signals drops dramatically. As the
input power of the large signals increases above the threshold,
the products of both small signals and large signals increase
linearly, i.e. $p_{out}$ proportional to $\sqrt{p_1 p_2}$. As the
input power keeps increasing,
eventually both small-signal and large-signal products become compressed.
We can refer the large signals in the simulation
to the un-correlated noise from the receivers, and
their product as the output fluctuations of the correlator,
as the output fluctuation of the correlator is indeed due
to the beating or mixing of un-correlated signals (Kraus, J. D. 1986).
The small signals in the simulation are regarded as
the correlated signals in the system. The simulation
defines three regimes of operation, namely, under-pumped, linear
and compressed. From experiments, our correlator modules have
a linear range for input power from -20 dBm to -12 dBm,
corresponding to output fluctuations with a root-mean-square
(RMS) from around 100 to 700 counts, including
the backend noise, as shown in Fig. 12.

As the correlators are operated in the linear regime,
it is seen that the correlator output fluctuations 
are proportional to 
$\sqrt{p_i p_j}$, where $p_{i, j}$ are the input power of
each baseline (Wrobel et al. 1999).
Since there is no total power detector at the correlator module inputs, 
the input power cannot be measured directly, but can be inferred from
the measured RMS of each baseline. In the compressed regime, the output RMS 
grows monotonically, although not linearly, as the input power increases. 
The RMS serves as an indicator of which regime we are operating in. The 
gain can then be adjusted to optimize the S/N.

\subsection{Spurious and Backend Noise}

An ideal double-balanced mixer used as a multiplier can suppress even harmonics
of both input signals and their products, i.e. $v_1^n$, $v_2^m$, and $v_1^n v_2^m$, 
where either n or m is even and $v_i (i= 1, 2)$ is the input voltage (Maas 1993).
From simulations, it has been found that with a double-balanced mixer,
the spurious terms such as $v_1^2$, $v_2^2$, $v_1^2 v_2^2$
are suppressed by over 40 dB below the desired product $v_1 v_2$.
Experiments show that residuals of these spurious terms can still be seen.
They can be further suppressed by the use of a phase switching/demodulation
process which allows the introduction of a DC blocking capacitor at the
input of the low frequency amplifier following the correlator module.
Phase switching is achieved by switching a suitable delay into
the LO path. However, as the LO signal is switched through different paths, 
a small amplitude modulation of the IF power level is introduced.
This modulation survives the demodulation process and appears as an 
offset in the $v_1^2 v_2^2$ term. To minimize offset, variable attenuators 
are used to adjust the 42 GHz LO power to the SHM to minimize the IF 
power modulation.

S/N is optimized by operating the correlator in a slightly compressed
regime, thereby ensuring that front-end noise dominates backend noise
from the DC amplifier and readout electronics.
Fig. 12 shows the output S/N estimated from
the Jupiter fringes as the input power were varied. 
Most of lag outputs shows an increase in S/N as the input power (infered
from the output RMS) increases, even when some of them were
driven into the compressed regime. 

A correlation interferometer where a multiplier is employed,
the minimum detectable temperature or sensitivity is 
\begin{equation}
\Delta T_{min} = \frac{T_{sys}}{\sqrt{2 \Delta \nu \tau}}
\end{equation}
where $T_{sys}$ is the system noise temperature,
$\Delta\nu$ is the bandwidth, and $\tau$ is the
integration time, assuming 100$\%$ efficiency. (Kraus, J. D. 1986).
From the correlator output fluctuations, we can estimate the noise contributions from each
part of the system, namely the front-end ($v_1^m$, $v_2^n$, and $v_1^m v_2^n$ terms,
where $m$ and $n$ are integers, $v_1$ and $v_2$ represent the large un-correlated noises.)
and backend (Fig. 11), assuming all noises are un-correlated, i.e. their variances 
are additive. Typical correlator output fluctuations measured under different 
conditions are listed in Table 2 to illustrate how the output noise might 
increase due to the spurious terms. As a result, noise contributions from 
the backend, as well as the spurious $v_1^m$, $v_2^n$ terms can be estimated. 
Currently these spurious terms and the backend noise reduce the S/N by 20$\%$.
An analysis of the system efficiency based on the observations
with the 7-element array can be found in Lin et al. (2009).

\subsection{Low-Frequency Amplifier}
For the low-frequency low-noise amplifier following the
lag correlator module, 
because of the high output impedance of the mixers,
low noise current amplifiers\footnote{OPA627AU}
were chosen 
to minimize the backend noise.
The bandwidth of the amplifier is limited in the feedback loop 
in order to reduce the output noise.
Due to the phase switching, the correlated signals are square waves,
at the phase switching frequency. A DC blocking
capacitor is used in front of the amplifier to remove any DC term from the
mixer output. As a result, the "DC" amplifier has a 3 dB passband from
0.1 to 9 KHz. The schematic of the DC amplifier is shown in Fig. 13.

\section{Readout Electronics}

The correlator readout circuit uses a
VFC plus a 24-bit counter 
as the ADC.
The VFC generates pulse sequences at a frequency
linearly proportional to the analog signal from the
DC amplifier output.
The up/down counter accumulates the pulses,
acting as a long-term integrator
as well as a phase-switching demodulator.
The up/down function is controlled by the
demodulation signal. 

In AMiBA correlator there are four layers of processes used to remove
systematics. The correlation process suppresses the uncorrelated receiver 
noise and gain fluctuations. The phase switching and demodulation process
reduces the offsets or gain drifts that do not have the characteristics of the
demodulation signals. One additional phase switching layer is implemented by inverting
the sign of the demodulation signals for every other integration. Every two data
points are then subtracted with respect to each other offline in the CCC.
From analysis, the correlator output spectra show a
white noise signature between $10^{-4}$ and 1 Hz, with an increase in 
power at lower frequency due to slow gain drifting (Nishioka et al. 2009). 
For the remaining offsets or false signals due to ground pick up, a 
two-patch scheme (Padin et al. 2002) is adopted. Observations are 
taken with, for our case, 3-minute tracking on the main field, and 
another 3-minute tracking on a trailing or leading field at the 
same declination, separated by few minutes in right ascension. 
The main and the trailing/leading fields then share the same azimuth-elevation 
track with identical ground contamination of the data.

At the end of an integration interval,
the contents of the counters are dumped into
shift registers and serially scanned out by the data acquisition
electronics. For the 13-element array, instead of the custom
readout ICs used previously in the 7-element array (Li et al.
2004), 
discrete VFC components\footnote{AD7741, Analog Devices} are used. The digital section
(counters/shift registers) is implemented in a field programmable
gate array (FPGA). The new VFC is synchronous, and has better
linearity than the previous readout ICs.
A schematic of the readout electronics with the timing
diagram of the control signals is shown in Fig. 14.

\section{Data Acquisition}

The data acquisition circuit is used to store data from the
readout circuits before they are transferred to the CCC via a direct memory access (DMA) process. 
Control signals, such as readout control signals and phase switching/demodulation signals are also generated here.
RAM blocks configured as $128 \times 24$ bit RAMs within the FPGAs are allocated to 
store the correlation and total power data. 
There are two steps involved in transferring
the data from the readout circuits to the CCC, scan and DMA. 
Each step is triggered by events originating from the Event Generator (EG) in the CCC.
The DMA process is ideal for transferring
large volumes of data but is more efficient if the data are
stored in consecutive addresses.

For the 7-element array, Walsh functions of 64 intervals per cycle are used as the phase switching signals. Currently there are 5 cycles in each integration
of 0.226 second. This corresponds to a fastest switching frequency of around 700 Hz. For the 13-element array, Walsh functions of 128 intervals per cycle are used. Since the corner frequency of the 1/f noise from the passive diode multiplier we use is low, phase switching at rather low frequencies is feasible.

\section{Control}
The operation of the AMiBA correlator is controlled by 
the CCC - an industrial grade PC running Linux. 
The CCC is equipped with 3 special cards -
Event Generator (EG), Australia Telescope Distributed Clock (ATDC),
and PCI Correlator Data Bus Interface (PCIIF) from ATNF
\footnote{http://www.atnf.csiro.au/technology/electronics/}. 
The function of the EG is to generate events with precise
timing. The PCIIF acts as an interface
between the data acquisition circuit and the CCC to receive the DMA data. 
The interface appears as a 256K byte block of memory on the
PCI bus while the
PCIIF also assigns memory addresses and several control
signals for the DMA process. Interrupts
from the EG are
relayed to the PCIIF for DMA timing.
The ATDC 
provides precise timing for all the correlator operations and
also generates an 8 MHz reference clock for digital processing
in the data acquisition and readout circuits. The ATDC can phase-lock
to both a 5(or 10) MHz sine wave and a 1 pulse-per-second (PPS) signal
from a GPS receiver and
is scheduled to synchronize the system clock of the CCC
periodically. An alternative is to synchronize the system clock of the CCC
to the GPS receiver via the network time protocol (NTP).
A block diagram of correlator control components and signals is
shown in Fig. 15.

\section{Testing and data processing}

To test the response of each baseline,
a W-band noise source is set up to translate between
the 2 receiver inputs. 
The data of lag sequence can be Fourier transformed to
obtain the bandpass response.
The bandpass shows a two-hump
gain response and some scatter from a linear phase response.  
From the bandpass response, we can estimate the effective
bandwidth $B_e$ as
\begin{equation}
B_e =  \frac{\left| \int W(f) df \right|^2}
{\int \left| W(f) \right|^2 df },
\end{equation}
where $W(f)$ is the complex bandpass response. 
The baselines of the 7-element array were estimated to 
have effective bandwidths ranging from 9 to 13 GHz
(Lin et al. 2009).
 
With the derived bandpass responses,
a transform matrix $K$ can be created so that
\begin{equation}
[K] [S] = [R],
\end{equation} 
where $S$ is the complex column vector representing the input spectrum, and
$R$ is the real column vector for the correlator output.
$K$ and $S$ include both positive frequency terms and their complex
conjugate at negative frequencies to yield real correlation products
(Li et al. 2004). 

With the correlation outputs from each baseline, we can invert $K$ and then
derive the spectra of the
input signals or visibilities $S$ via
\begin{equation}
[S] = [K]^{-1} [R].
\end{equation}
However, since the 4 lag outputs only give us 2 independent
complex data points in the frequency spectrum, $K$ is rather singular
due to degeneracy. The Singular Value Decomposition (SVD) method can be used to
invert $K$.
Once $K^{-1}$ is obtained, the spectra of $n$ $(n > 2)$ channels can be
obtained. Eventually we have to consolidate those $n$ channels into 2 bands.
Another approach would be to integrate the complex response matrix 
$K$ in frequency into two bands first. 
The integrated matrix $\bar{K}$ can
be easily inverted, and the 2-band visibility output can be written as $[\bar{S}]=[\bar{K}]^{-1}[R]$. 

From simulations, Lin et al. (2009) showed that given an accurate 
and high spectral resolution measurement of $K$, the visibility can be recovered regardless of the source offset from the phase center. On the other hand, inaccurate estimate of $K$ results in errors in the recovered visibility. The error 
varies with the source offset and can be calibrated at any given offset 
by observing a point source at the same offset from the phase center. 
The calibration is strictly valid only in a small region (a small fraction of the 
synthesized beam) around the location at which the calibration source is observed. 
At other locations within the field of view, there are gain and phase errors set 
by uncertainties in the lag-to-visibility transformation that affect the 
quality of the image reconstruction, although further calibration should 
reduce these errors. However, further simulations show that applying the calibration 
to the entire field of view only contributes about $2\%$ in RMS to the recovered 
point source flux after combining data from all baselines (Lin et al. 2008). 

Although using a translating noise source at the receiver inputs
to create artificial fringes is a useful method to measure the
transfer function of the correlator, it is hard to run the test inline
with observations. 
The spectral resolution obtained with our current setup was not sufficient.
This approach becomes more tedious as the number of
baselines increases and we would still need to use the planet data
to calibrate the gain and phase of each baseline. 
An alternative would be to derive the response matrix 
$K$ in two bands, assuming nominal responses of an ideal lag correlator. 
Data from tracking of strong point sources such as Jupiter, Saturn,
or Mars, performed regularly at an interval of approximately 3 hours during the observations, can be used for calibration. All data analysis and calibrations
were restricted to observations made at the phase center at the moment. 
Fig. 16 shows the images of Jupiter from visibilities with and without calibration. The calibrator was another set of data of Jupiter taken several minutes apart. The uncalibrated visibilities suffer from errors in the 
transformation as mentioned above. Forming an image directly results in very strong cancellation. After calibration, the visibilities add up coherently and form a strong point source at the phase center. The images have been deconvolved using the CLEAN algorithm (Hogbom 1974) and plotted with the same dynamical range.
More details about data processing can be found in Wu et al. (2009)
and Lin et al. (2009).

\section{Conclusion}
From observations, the AMiBA correlator has proven to have the sensitivity
required for CMB detection. The analog multipliers used in the
lag correlators provide the wide bandwidth required for high sensitivity.
The inherent noise rejection of the interferometer
is also very beneficial. Compared with
the filterbank scheme used by CBI (Padin et al. 2002), 
the lag correlator design is simpler
and more compact. Large bandwidth with a small number of
lags does not present a significant challenge for the lag-to-visibility
transformation after proper calibrations. At the moment,
the effective bandwidth we can achieve is limited by
bandpass variations due to wideband impedance matching. However, 
a significant portion ($60\%$) of the nominal bandwidth has been achieved.
 To improve on the bandpass variations, it is
 possible 
to operate the correlator at a higher IF, since the response
 of the analog multipliers is not limited to 
frequencies of few GHz. Thus, with a similar 
fractional bandwidth, a larger bandwidth could be 
achieved. 
Wideband complex correlators in conjunction with bandpass 
filters can 
also be considered. For AMiBA, complex correlators 
with a bandwidth of 
8 GHz are suitable. By interleaving a number of them in frequency, a large bandwidth 
is feasible. 
The filters would also improve the isolation between channels.
Similar analog interferometric systems can be constructed
for high sensitivity, high angular resolution, and 
moderate frequency resolution observations.

\begin{table}[H] \caption{AMiBA 4-Lag Correlator Specifications}
\begin{center}
\begin{tabular}{|l|l|}
\hline
Input Frequency Range & 2 - 18 GHz \\
\hline
Responsivity & 80 $V_{rms}$/W minimum\\
\hline
Responsivity Variation vs. frequency & $<$ 3 $dB_{pp}$\\
\hline
Phase Response & $<$ 30 degree peak-to-peak deviation from linear\\
\hline
Delay Increment Accuracy & 25.0 +/- 2.5 ps \\
\hline
Input 1 dB Compression Point & $>$ -5.0 dBm\\
\hline
Squared Term Contribution & $<$ 5 $\%$ \\
\hline
Output Impedance & $<$ 100 k$\Omega$ \\
\hline
\end{tabular}
\end{center}
\end{table}

\begin{table}[H] \caption{
A typical result of the correlator output fluctuations or RMS, measured under 4
conditions. When both receivers were off, the output noise could be attributed
to the backend noise from the DC amplifier and the ADC. When only one antenna
was on, e.g. Ant 2, the output of the multiplier would include various terms from
$V_1$. The situation was similar when only Ant 3 was on. For an ideal multiplier,
the $v_1^m$ and $v_2^n$ terms are considered spurious. From the measurements, these
terms also contributed to the output noise so that correlator output RMS
increased. Assuming that noise from each term can be summed in quadrature, we can
estimate the output fluctuation when there is no $v_1^m$ or $v_2^n$ terms
at the output and no backend noise either, so as to determine how much the
S/N has been degraded.}
\begin{center}
\begin{tabular}{| l | l | l | l | l | l | l |}
\hline
Measurement & 1 & 2 & 3 & 4 & &\\
\hline
Ant 2 ($v_1$) & off & off & on & on & &\\
\hline
Ant 3 ($v_2$) & off & on & off & on & &\\
\hline
YY lag 3 RMS (counts)& 95 & 783 & 805 & 1850 & 1120 & 1473\\
\hline
noise & backend & $v_1^m$ terms, & $v_2^n$ terms, & overall & $v_1^m$ terms, & $v_1^m v_2^n$ terms\\
composition &  & backend & backend &  & $v_2^n$ terms, & \\
   & &  & & & backend & \\
\hline
\end{tabular}
\end{center}
\end{table}

\begin{figure}[H]
  \begin{center}
  \includegraphics[width=4.8 in]{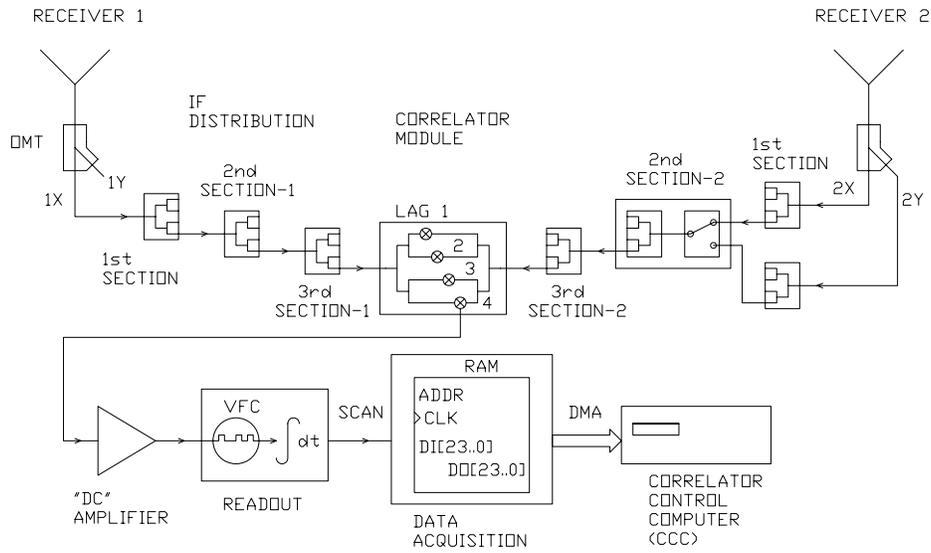}
  \caption{Block diagram of the AMiBA correlator. Signal flow for one
baseline (1X2X) is presented. Following the correlator module,
the signal flow for a particular lag (lag 4) is depicted.\label{fig:rxscheme}}
  \end{center}
\end{figure}

\begin{figure}[H]
  \begin{center}
  \includegraphics[width= 6.4 in]{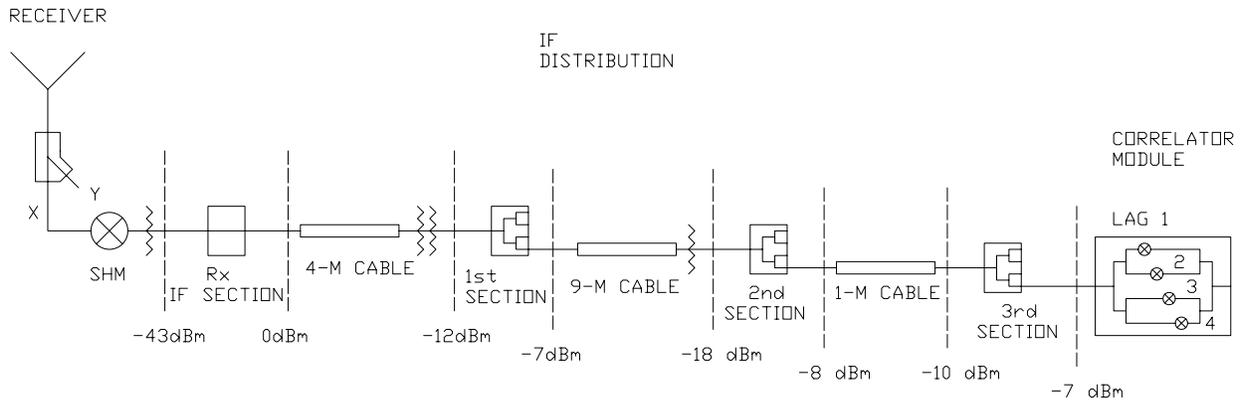}
  \caption{IF power profile from the receiver output to
correlator module input (prior to 2009).\label{fig:rxscheme}}
  \end{center}
\end{figure}

\begin{figure}[H]
  \begin{center}
  \includegraphics[width = 6.4 in]{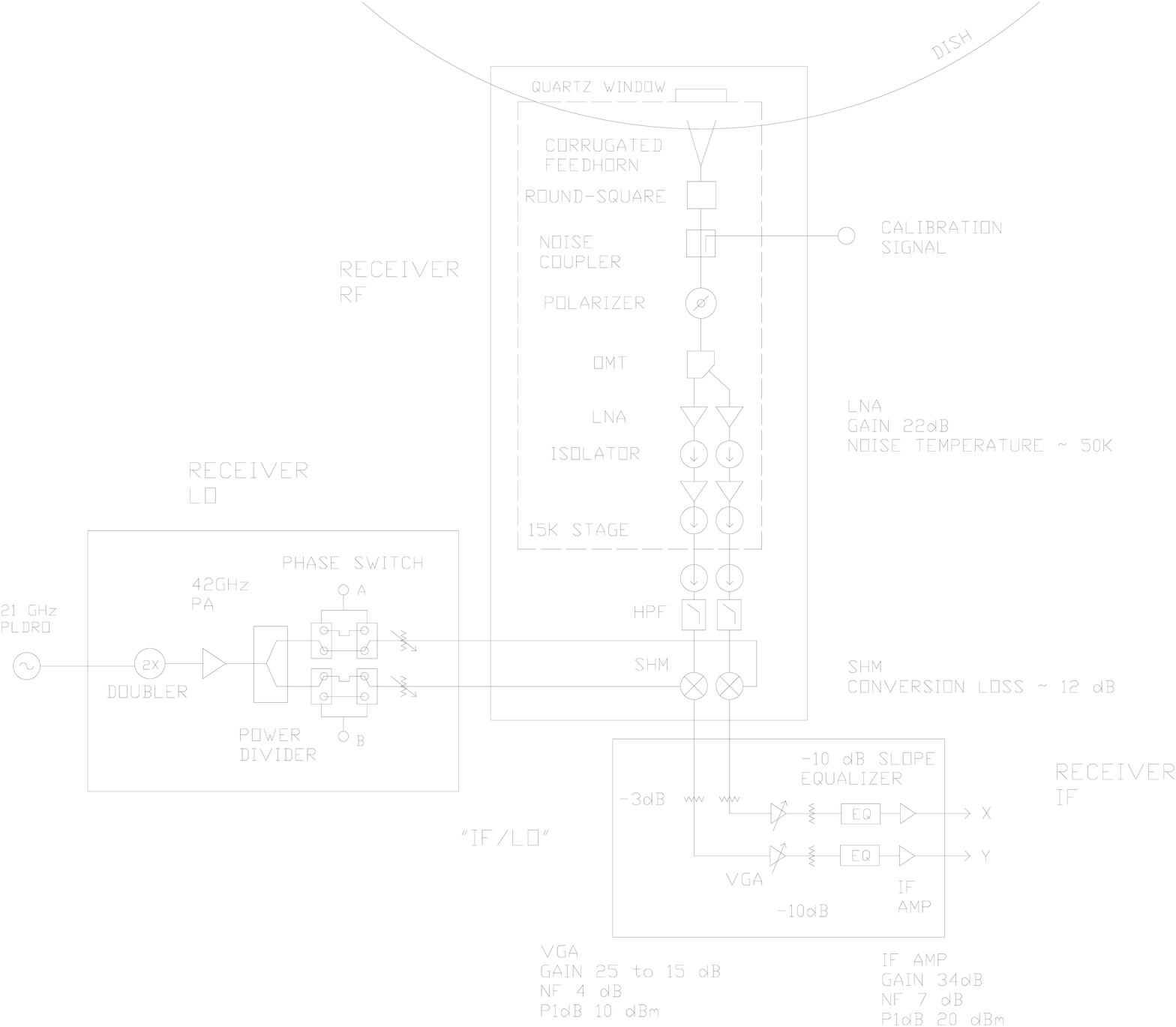}
  \caption{Schematic of the AMiBA receiver, including the
RF, LO, and IF sections.
\label{fig:rxscheme}}
  \end{center}
\end{figure}

\begin{figure}[H]
  \begin{center}
  \includegraphics[width=6.4 in]{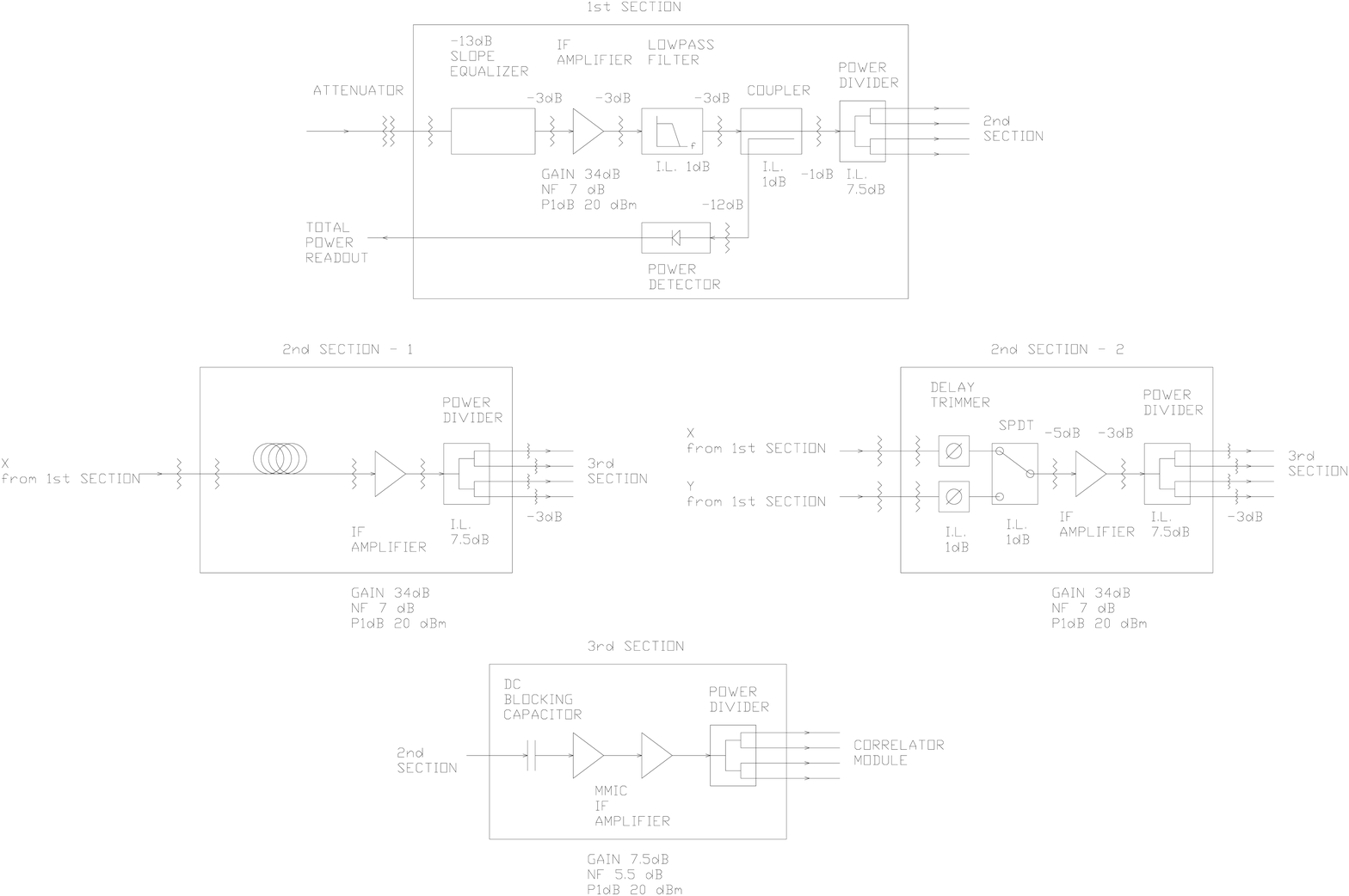}
  \caption{Schematic of the correlator IF sections. The gain, insertion loss
(I.L.), noise figure (N.F.), and output 1 dB compression point
(P1dB) of each component are listed where applicable.
\label{fig:rxscheme}}
  \end{center}
\end{figure}

\begin{figure}[H]
\begin{center}
\includegraphics[width=4.8 in]{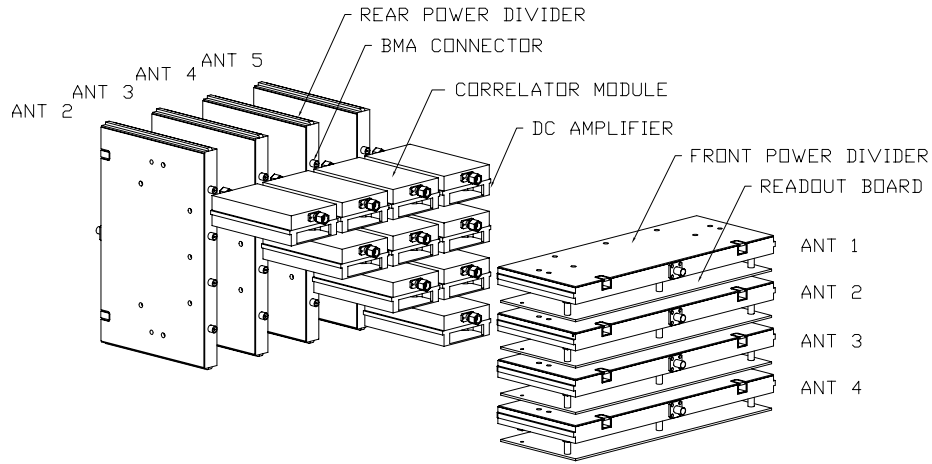}
\caption{Modular layout of the correlator 3rd section power dividers,
correlator modules, DC amplifiers, and readout boards. The central
portion consists of an array of lag correlators and DC amplifiers. The correlators
are fed by power dividers at both ends. Readout boards underneath the
horizonal power dividers are responsible for VFC ADC and integration.}
\end{center}
\end{figure}

\begin{figure}[H]
  \begin{center}
  \includegraphics[width=3.2 in]{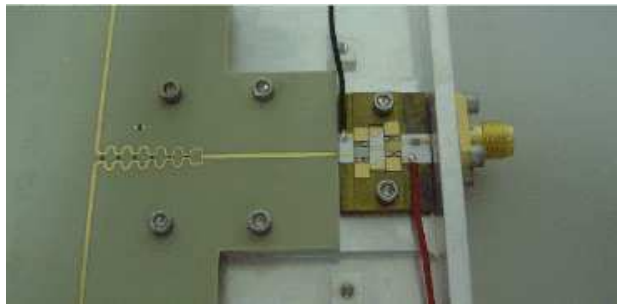}
  \caption{Picture of the MMIC IF amplifiers and
power dividers used in the 3rd section of correlator IF.
Bias circuit boards are not shown.
\label{fig:rxscheme}}
  \end{center}
\end{figure}

\begin{figure}[H]
  \begin{center}
  \includegraphics[width=4.8 in]{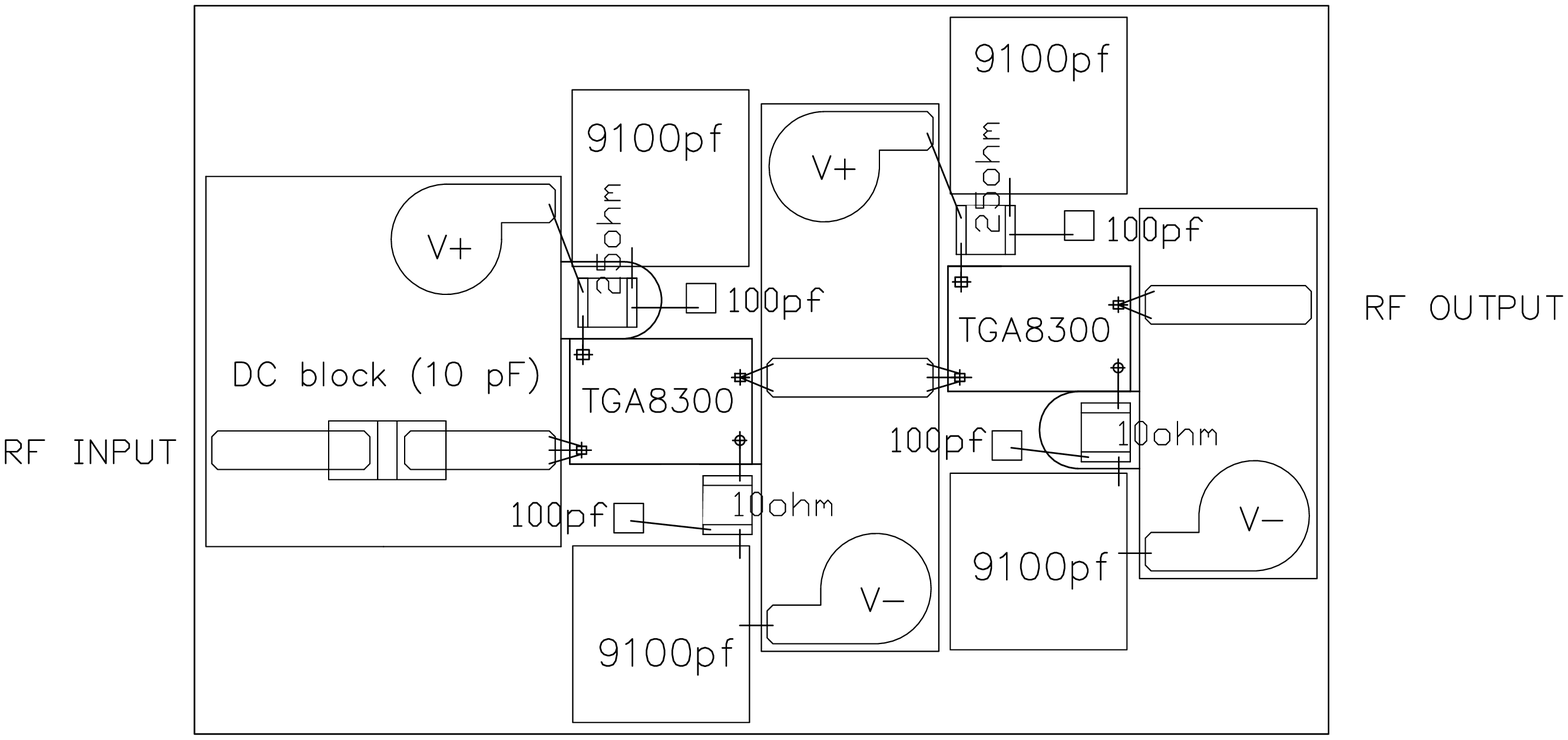}
  \caption{Assembly diagram for
the IF amplifiers used in the 3rd section of the correlator IF as
recommended by the manufacturer, except the 100 pF bypass
capacitors used to remove the oscillation at high frequencies.\label{fig:rxscheme}}
  \end{center}
\end{figure}

\begin{figure}[H]
  \begin{center}
  \includegraphics[width=4.26 in]{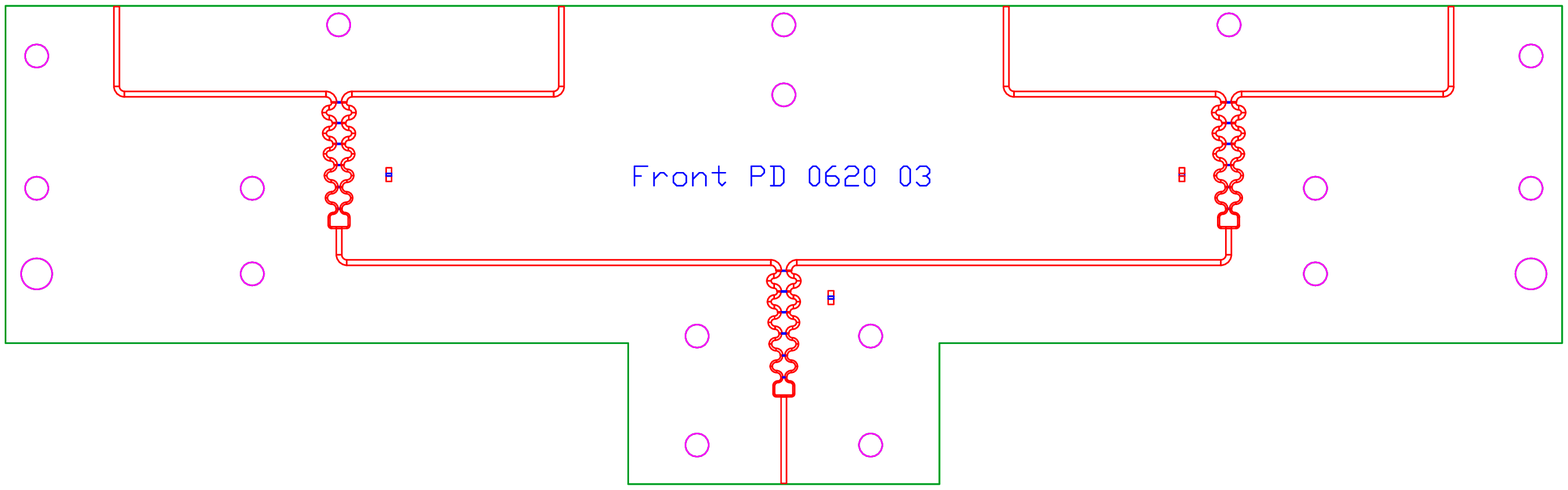}
  \includegraphics[width=3.2 in]{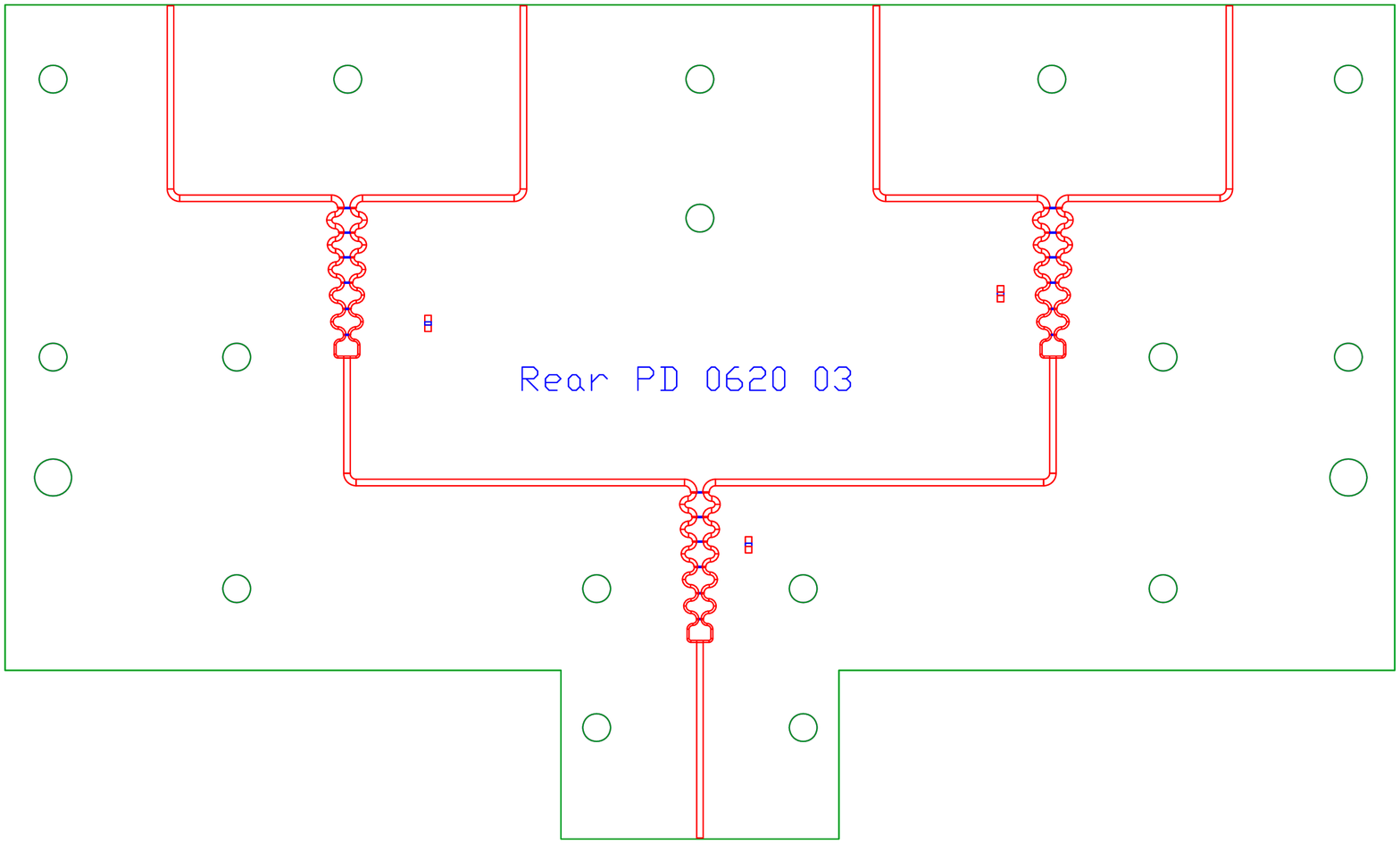}
  \caption{Layouts of the custom 4-way power dividers\label{fig:rxscheme}}
  \end{center}
\end{figure}


\begin{figure}[H]
  \begin{center}
  \includegraphics[width=4.8 in]{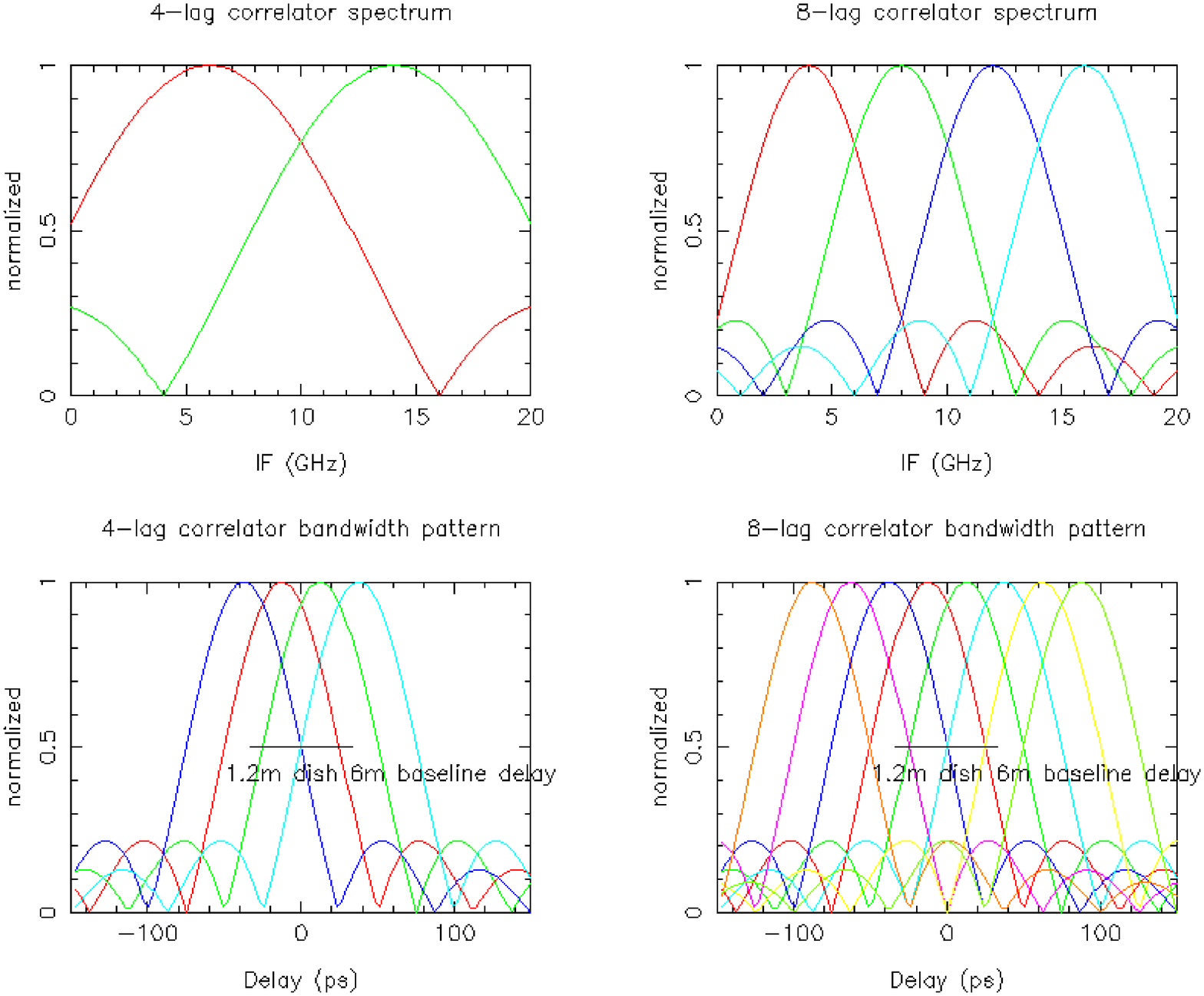}
  \caption{{\it Upper Left}: Recovered spectrum for a 4-lag correlator with CW input
signals at the center of 2 channels. {\it Upper Right}: Spectrum for a 8-lag correlator
{\it Bottom Left}: Bandwidth pattern for a 4-lag correlator and delay range for
1.2m dishes with 6m baseline {\it Bottom Right}: Bandwidth pattern for a 8-lag
correlator\label{fig:rxscheme}}
  \end{center}
\end{figure}

\begin{figure}[H]
\begin{center}
\includegraphics[width=3.2 in]{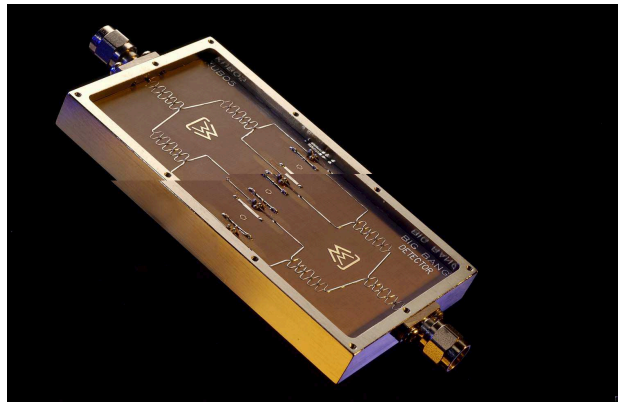}
\caption{The 4-lag correlator module manufactured by Marki Microwave Inc. 
The power dividers and multipliers can be seen.}
\end{center}
\end{figure}

\begin{figure}[H]
  \begin{center}
  \includegraphics[width=4.0 in]{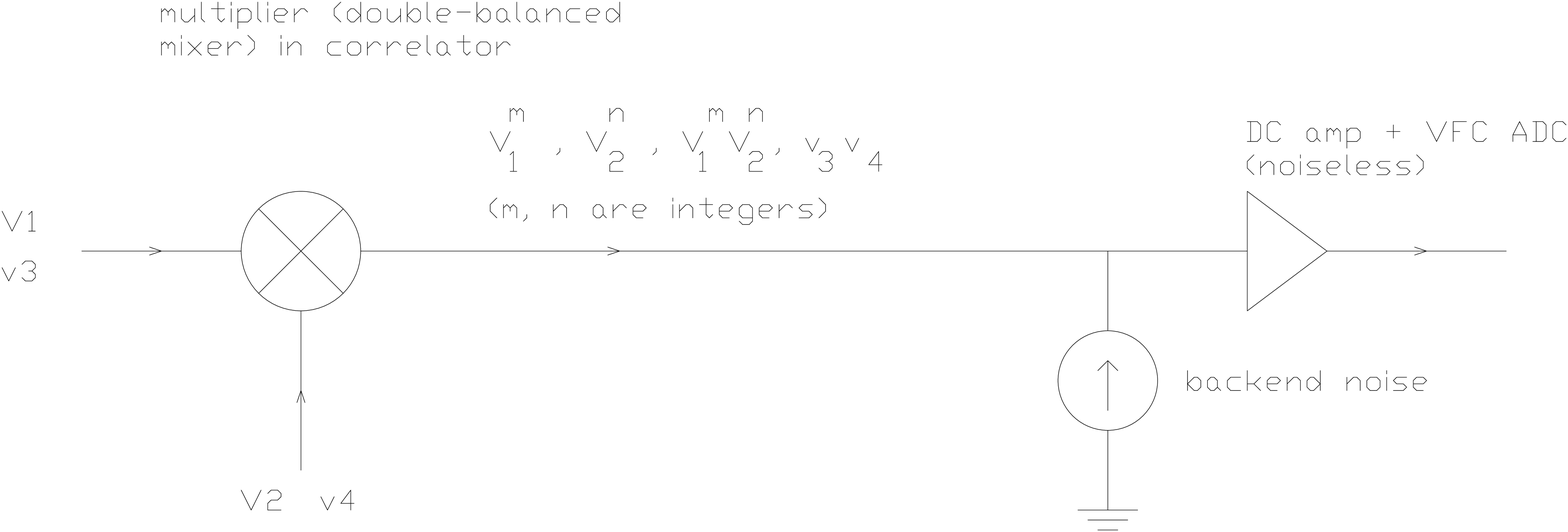}  
   \includegraphics[width=4.0 in]{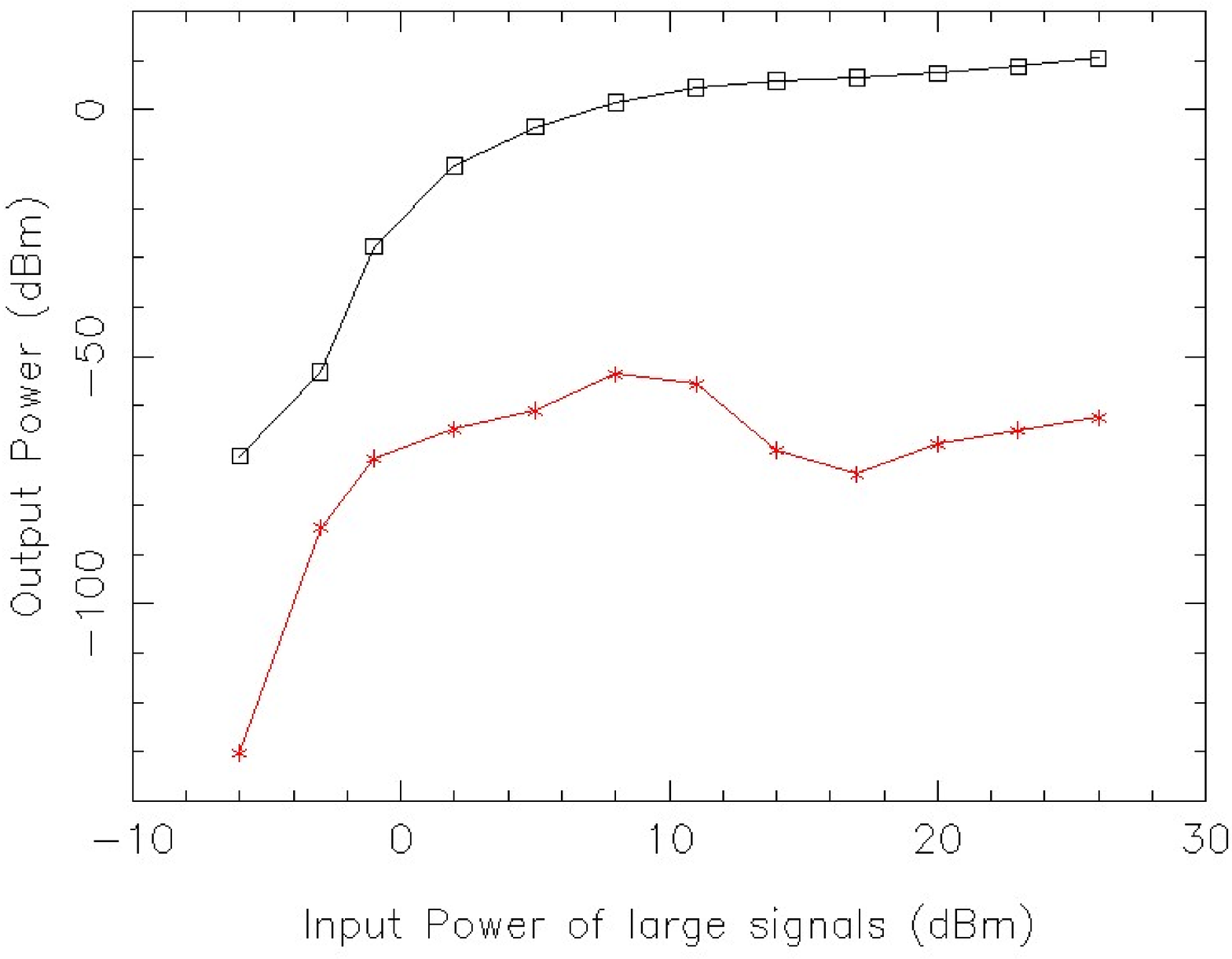}

  \caption{{\it Upper}:
With $v_1$ and $v_2$ as the large un-correlated noises, 
the output from a multiplier can be represented as the 
summation of various possible terms $v_1^m$, 
$v_2^n$, and $v_1^m v_2^n$, where $m$ and $n$ are integers.
Each output terms can contribute to 
correlator output fluctuations. 
On the other hand, $v_3$ and $v_4$ present the small correlated 
signals, and $v_3 v_4$ is the expected product. A current noise 
source is used to represent the backend noise from the DC amplifier
and the VFC ADC at the input of the DC amplifier.  
{\it Bottom}: 
In the simulation of a double-balanced mixer with 4 tones, when the 
large signals have power below a certain level (-2 dBm), the product 
of 2 small signals (red curve) drops dramatically. As the input power 
of the large signals increase above the threshold, products of both small 
signals and large signals (black curve) increase linearly. As the input 
power keeps increasing (above 8 dBm), eventually both small-signal 
and large-signal products get compressed. Since a standard diode model
is used during the simulations, the input power levels of the large 
signals where the diodes are sufficiently pumped or compressed are
different from our measurements.
}
  \end{center}
\end{figure}

\begin{figure}[H]
  \begin{center}
  \includegraphics[width=3.2 in]{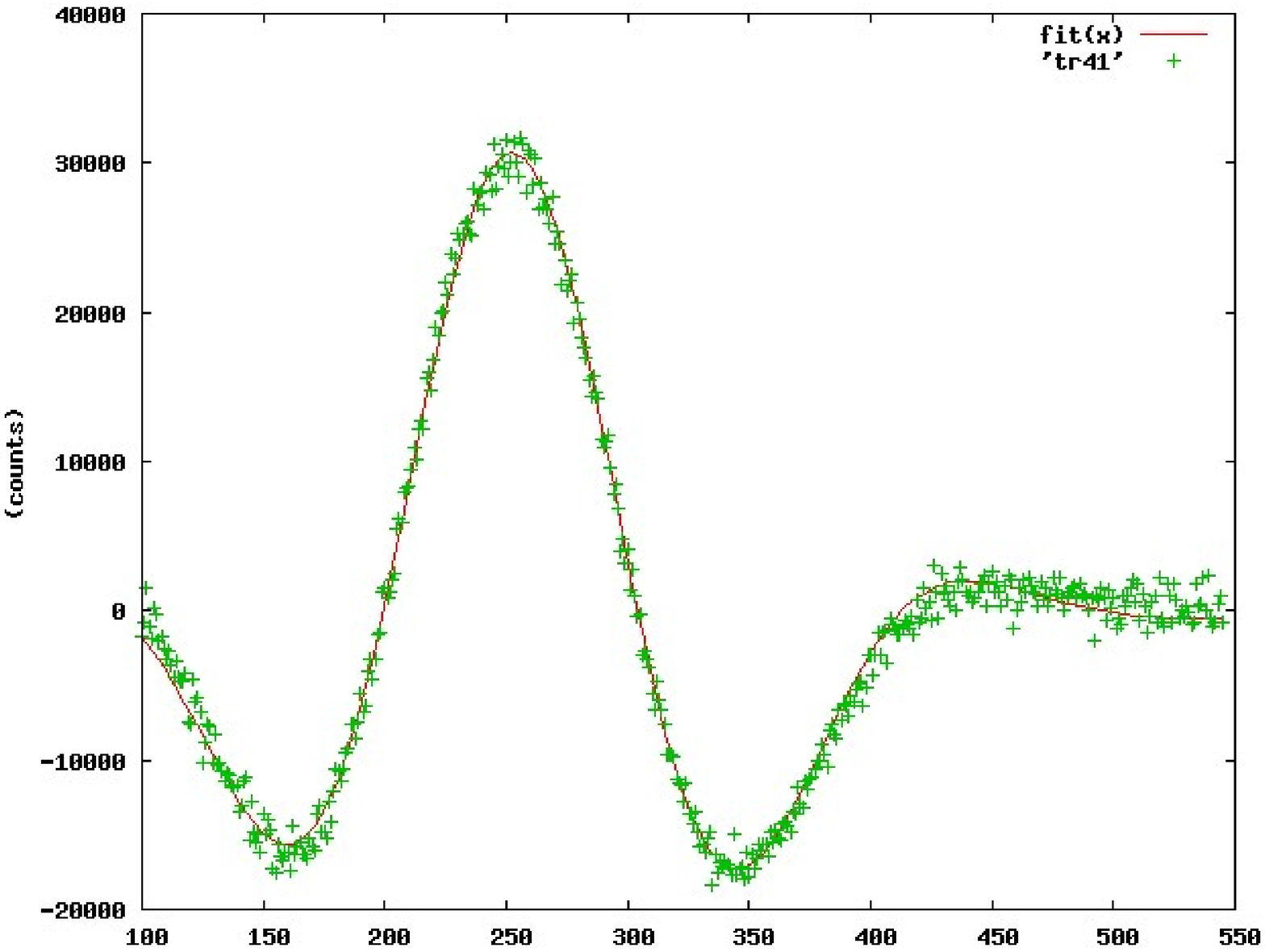}
  \includegraphics[width=3.2 in]{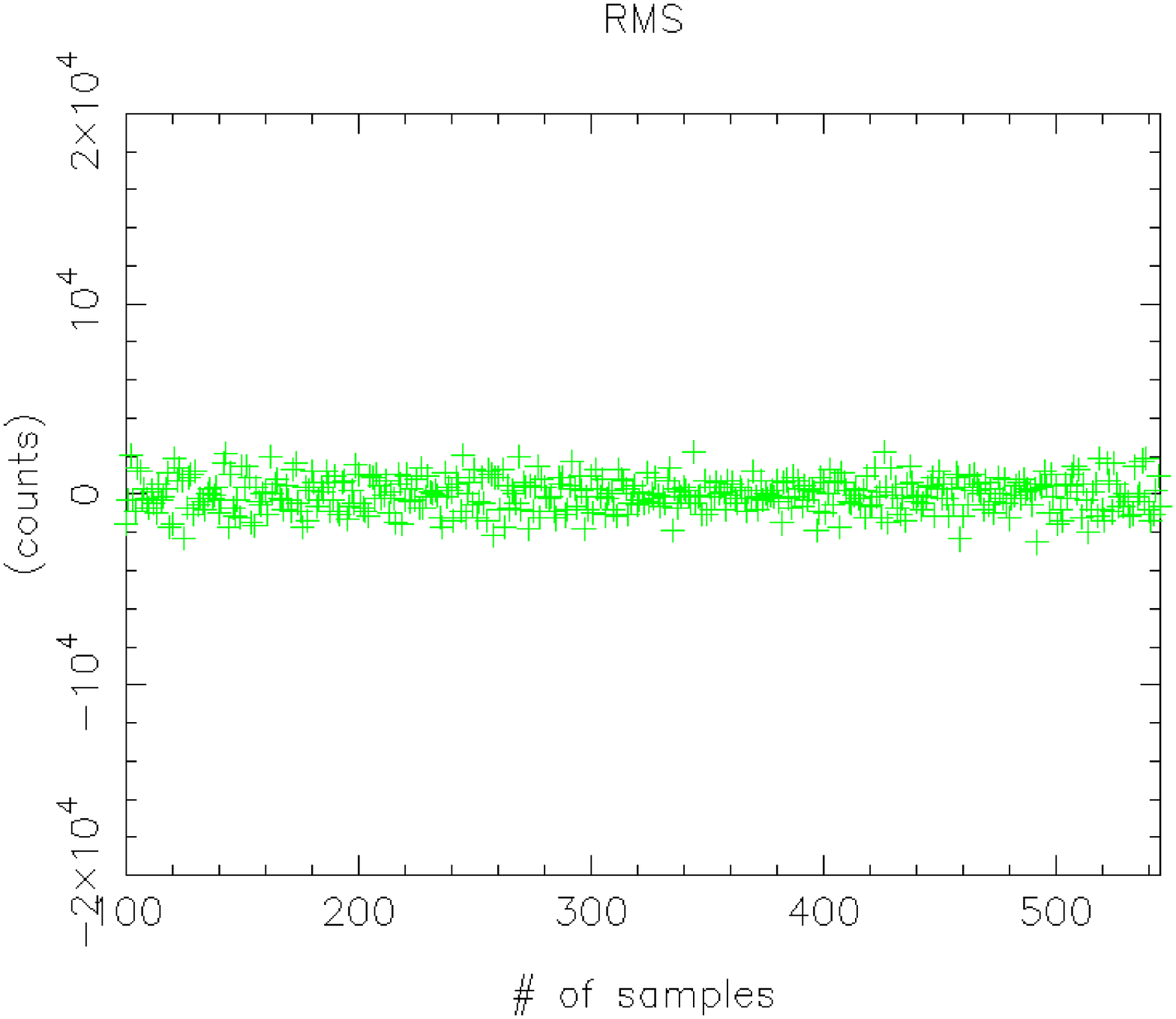}
  \includegraphics[width=3.2 in]{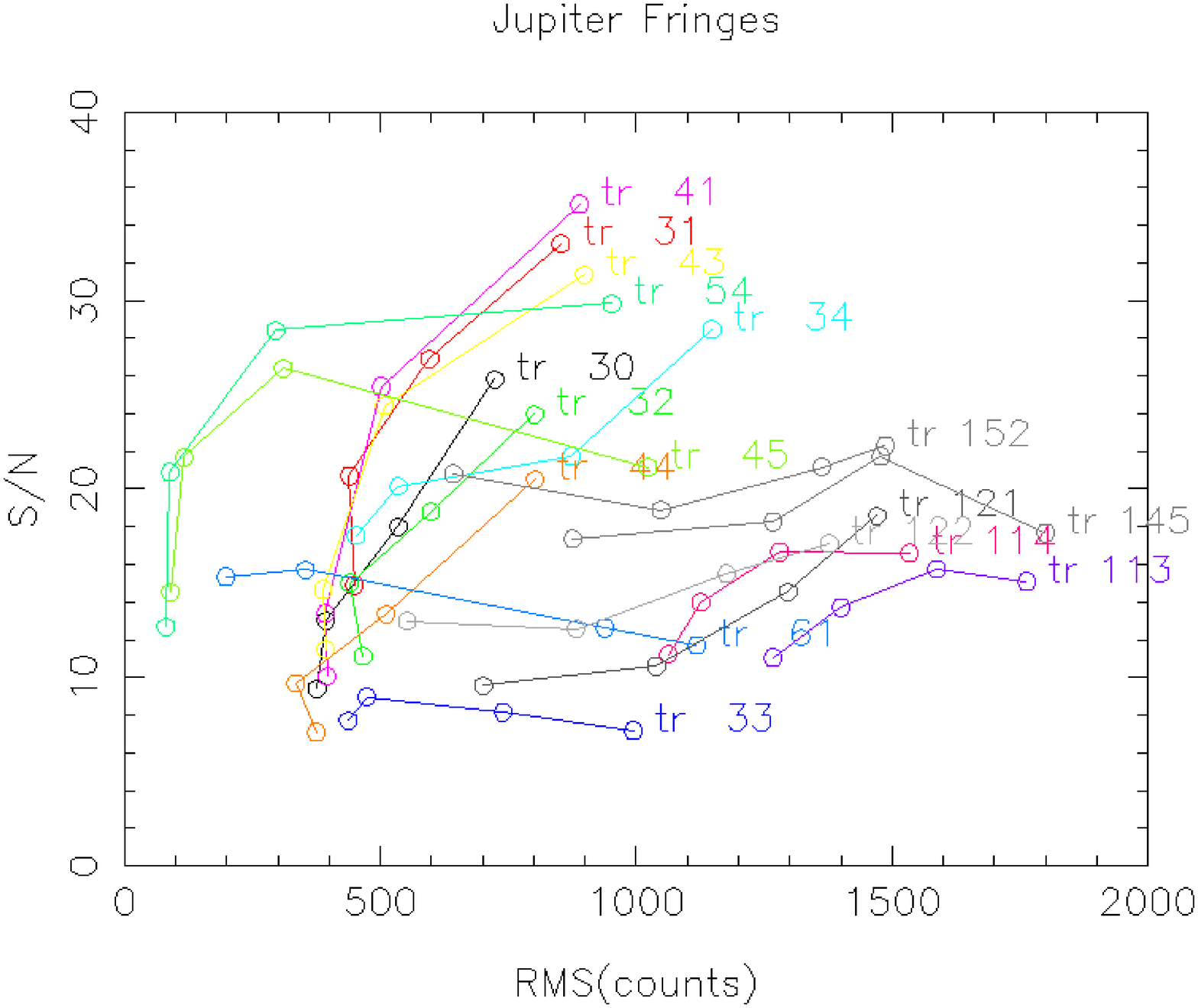}
  \includegraphics[width=3.2 in]{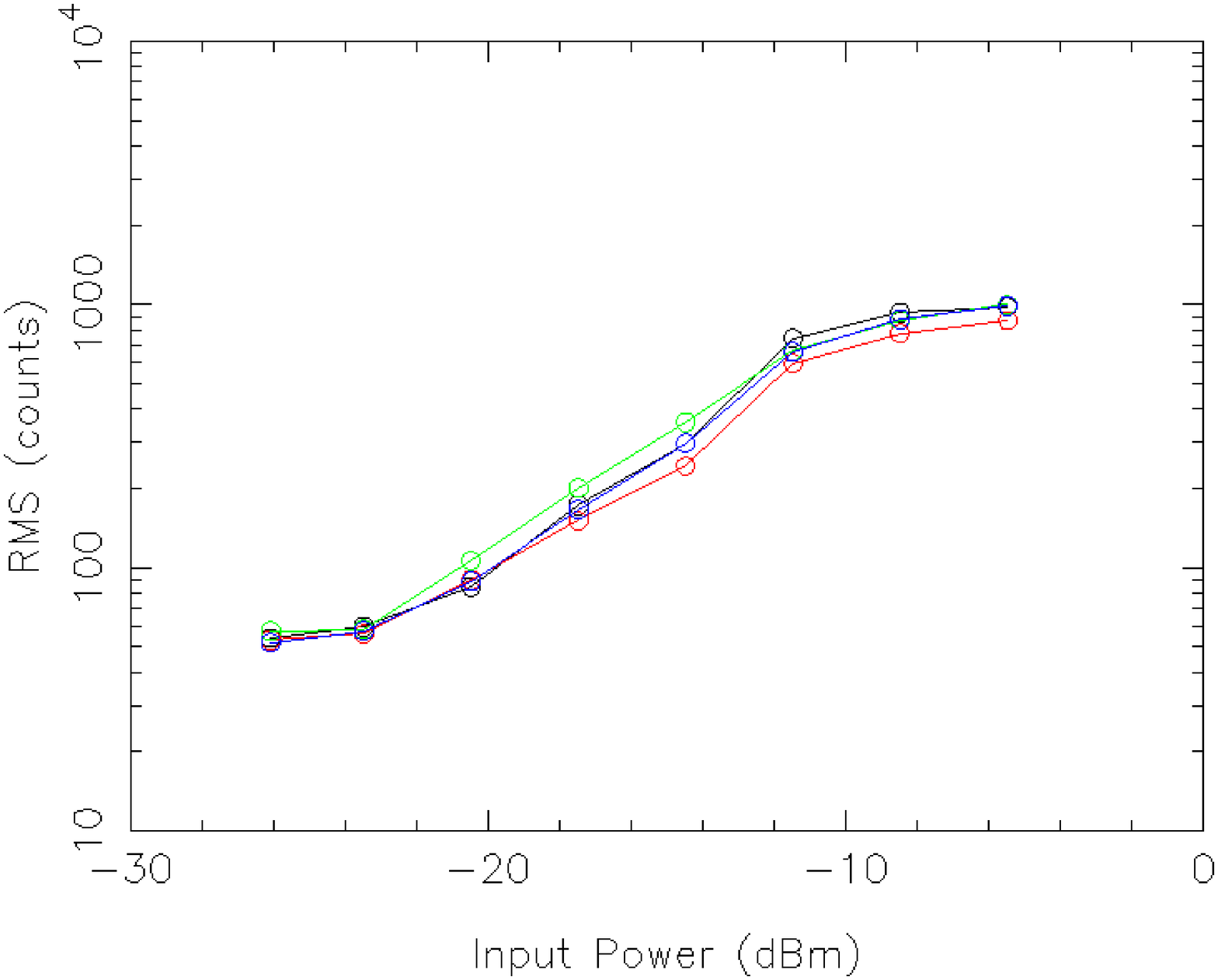}
  \caption{To determine the IF power for optimum S/N at correlator outputs, drift scans of
Jupiter were taken with different input power levels to the correlator. The input
power were varied by adjusting the variable gain amplifiers along the IF paths. The
input S/N was fixed and could be referred to the values at the receiver inputs.
Output from the 4-lag correlator module, usually referred
to as the lag output, can be fitted to estimate the signal strength. 
During data processing, each of 168 lag outputs of the 7-element
array is designated with a trace number. {\it Upper Left}: One lag output (trace \# 41, 
labelled as tr41) is plotted (green crosses) and curve-fitted (red line), in terms of counts 
after the VFC ADC. {\it Upper Right}: High-pass filtering is applied to the data to remove the low-frequency
terms and the signal. The noise is estimated from the RMS of the remaining fluctuations
(green crosses). The responses of the lag correlators vary. {\it Lower Left}: the S/N of several lag outputs (circles), 
are plotted against the RMS values. Lines are drawn through data points
of each output, and different colors are used for distinction. {\it Lower Right}:
For 4 lag outputs of one baseline, output fluctuations (RMS in counts),
including the backend noise, are plotted (circles) as a function of the input power.}
  \end{center}
\end{figure}

\begin{figure}[H]
  \begin{center}
  \includegraphics[width=4.8 in]{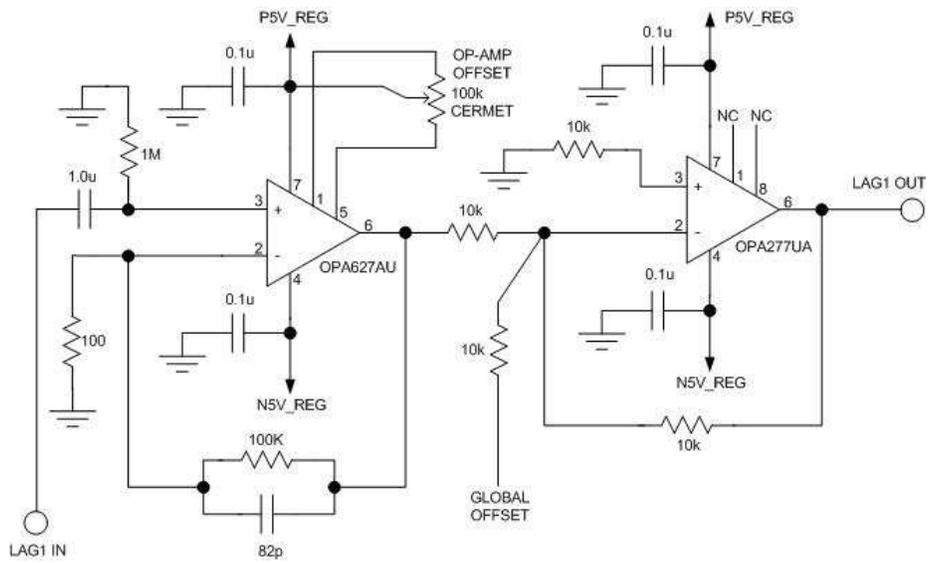}
  \caption{Schematic of the low frequency or "DC" amplifier\label{fig:rxscheme}}
  \end{center}
\end{figure}

\begin{figure}[H]
  \begin{center}
  \includegraphics[width=6.4 in]{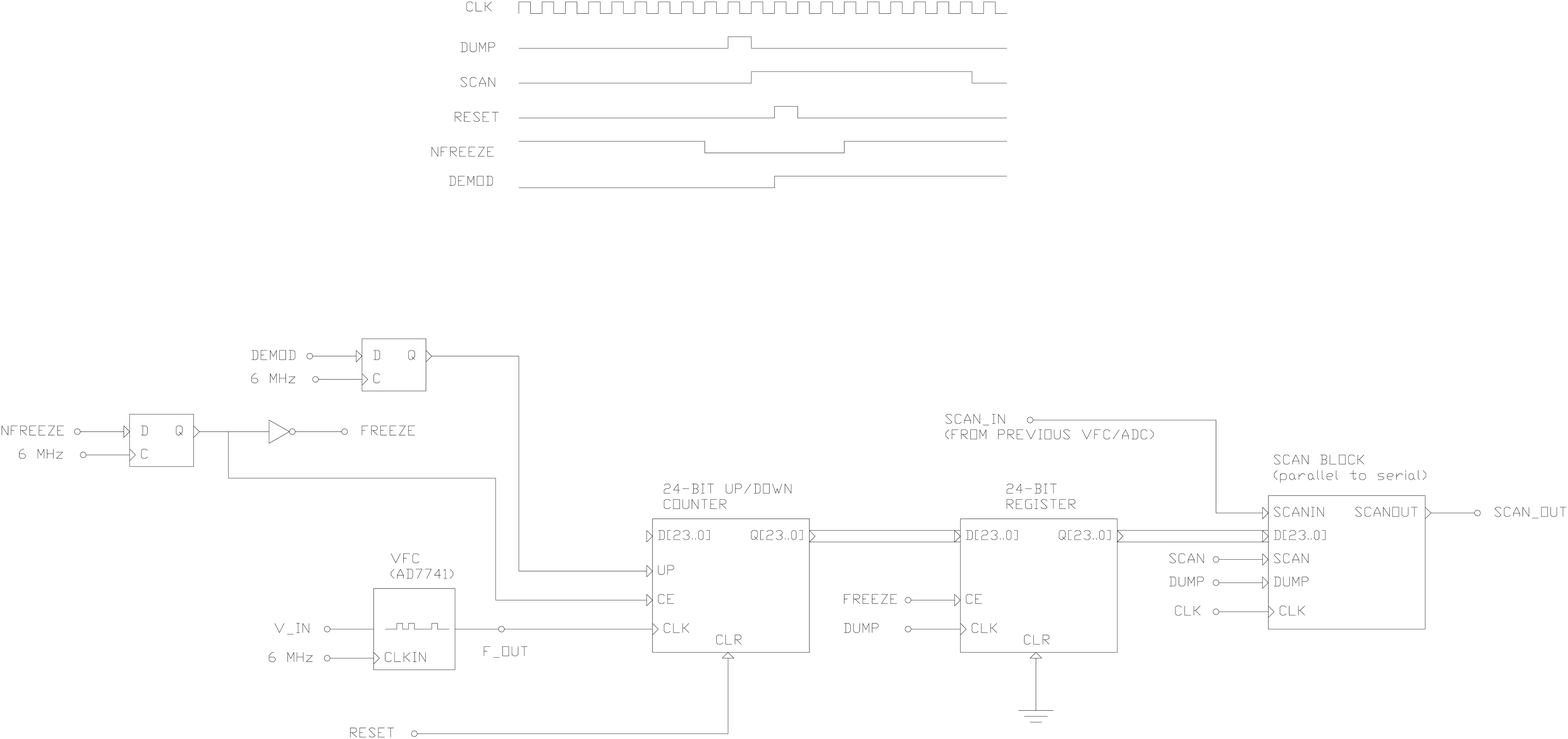}
  \includegraphics[width=6.4 in]{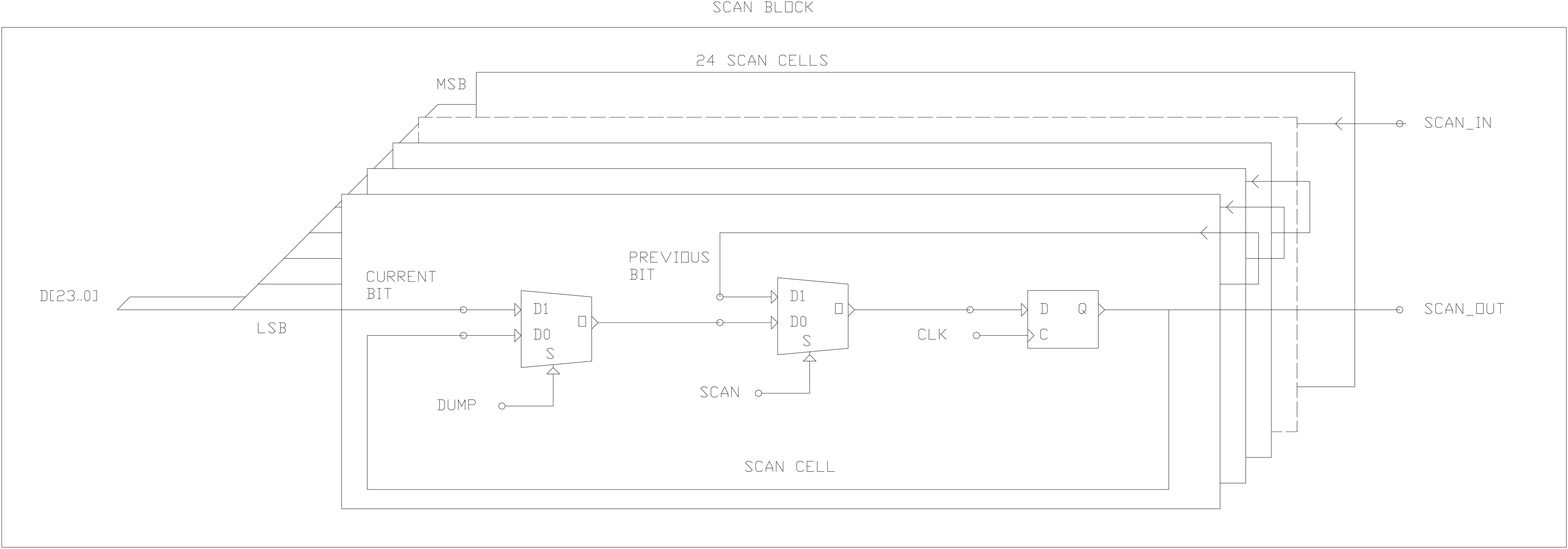}
  \caption{{\it Upper}: The timing diagram of the control signals and 
the schematic of the VFC/counter ADC
used in the correlator for the 13-element array.
Signals (dump, scan, reset, and demod - demodulation)
are applied to the counter only when counting is stopped ("freezed").
"Freeze" needs to be buffered before being applied to the counter to avoid
glitches. Since AD7741 is a synchronous VFC, i.e. the output pulse 
is initiated by the edge of the clock (CLKIN),
"freeze" signal is buffered with a flip-flop triggered by the clock
of the VFC.
{\it Bottom}: Schematic of the scan block used to transform 24-bit data
into series of bits for the scan-out process.
\label{fig:rxscheme}}
  \end{center}
\end{figure}

\begin{figure}[H]
  \begin{center}
  \includegraphics[width=6.4 in]{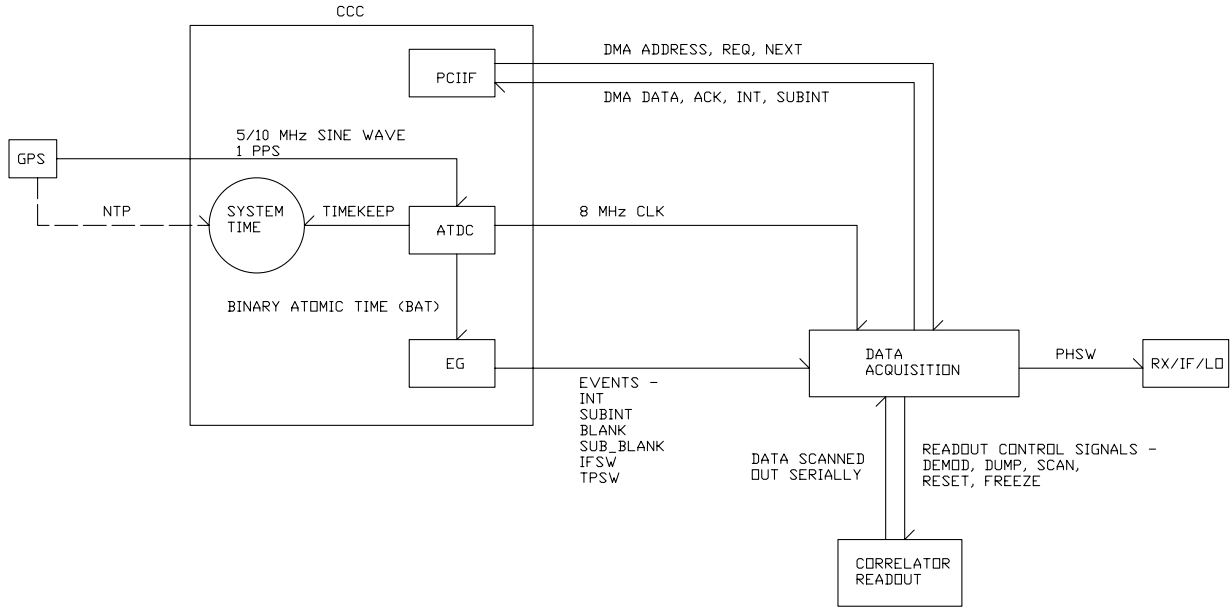}
  \caption{AMiBA correlator control block diagram.\label{fig:rxscheme}}
  \end{center}
\end{figure}

\begin{figure}[H]
\begin{center}
\includegraphics[width=3.2 in]{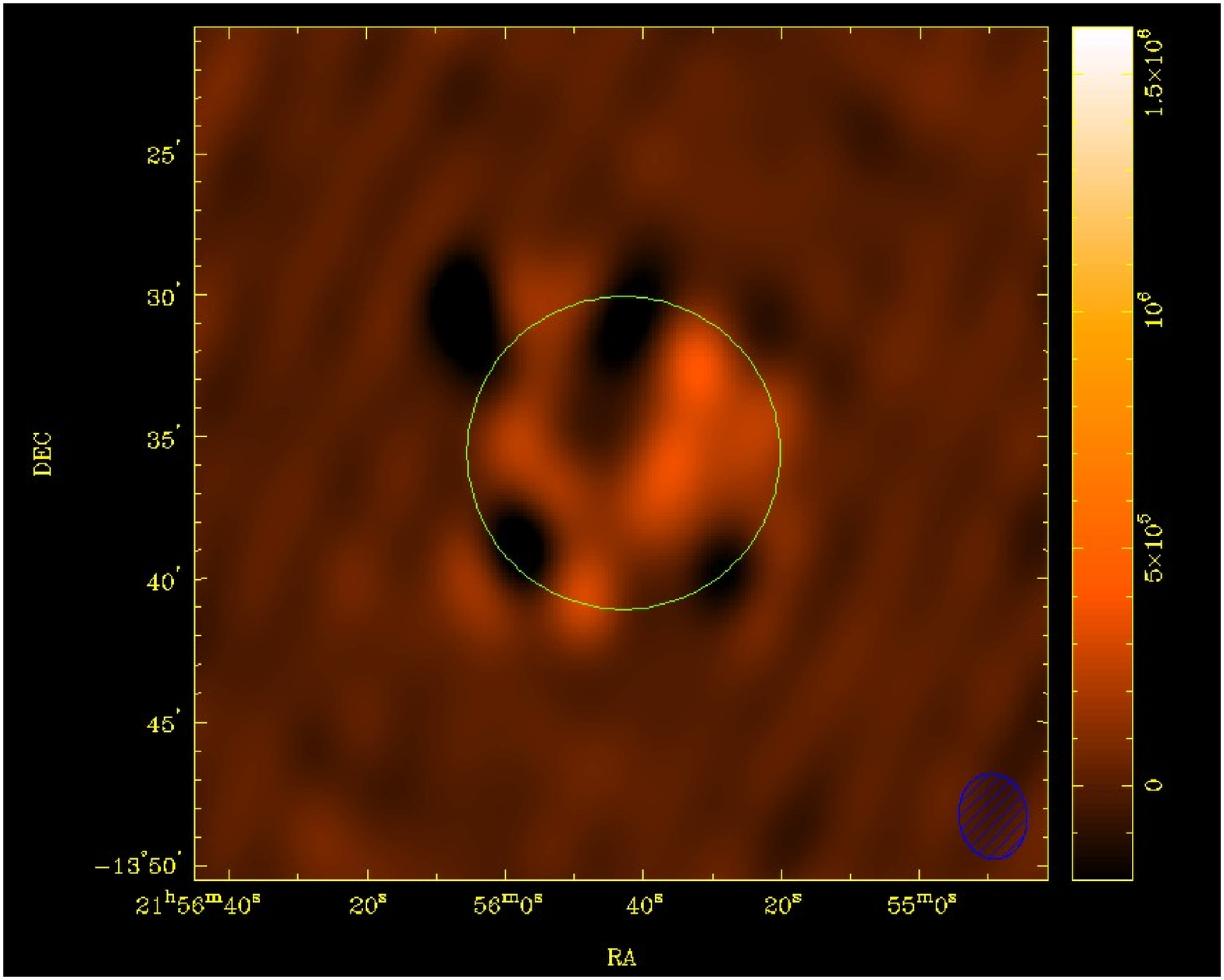}
 \includegraphics[width=3.2 in]{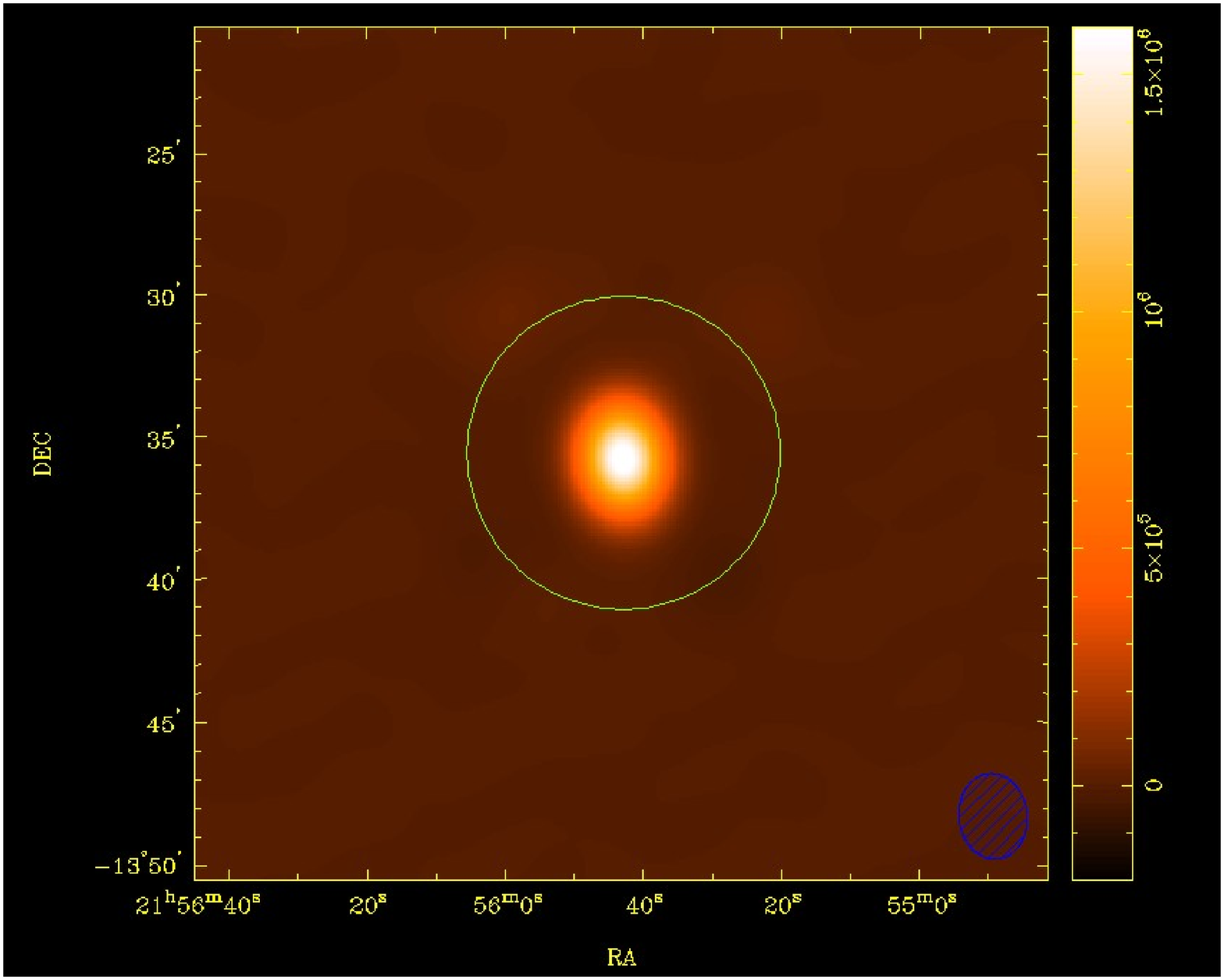}
 
    \caption{
The cleaned images of Jupiter. {\it Left}: The image was formed with uncalibrated visibilities directly transformed from correlator outputs. {\it Right}: The constituent visibilities of the image have been calibrated by another set of Jupiter data. Both images are plotted with the same residual noise after CLEAN.
The green circle indicates the FWHM of the primary beam, and the blue 
shaded ellipse at the bottom right corner represents the synthesized beam
of the 7-element array in the compact configuration.
    \label{fig:jupiter}
    }
\end{center}
\end{figure}

{\bf Acknowledgments.} 
We thank the administrative staff for their support over the years. We thank A. Harris for useful discussions and notes. We thank the Ministry of Education, the National Science Council, and the Academia Sinica for their support of this project. We thank the NOAA for accomodating
the AMiBA project on their site on Mauna Loa.  We thank the Hawaiian people for allowing astronomers to work on their mountains in order to study the Universe.

\end{document}